\UseRawInputEncoding
\documentclass[aps,pra,twocolumn,floatfix,superscriptaddress,showpacs,Superscript citations]{revtex4-1}
\usepackage{appendix}
\usepackage{graphicx}
\usepackage{amsfonts}
\usepackage{amsmath}
\usepackage{amssymb}
\usepackage{bm}
\usepackage{color}
\usepackage[hypertex]{hyperref}%
\providecommand{\U}[1]{\protect\rule{.1in}{.1in}}

\begin{document}

   \title{Anomalous supercurrent modulated by interfacial magnetizations in Josephson junctions with ferromagnetic bilayers}

   \author{Hao Meng}
   \affiliation{Shenzhen Institute for Quantum Science and Engineering (SIQSE), Southern University of Science and Technology, Shenzhen 518055, China}
   \affiliation{School of Physics and Telecommunication Engineering, Shaanxi University of Technology, Hanzhong 723001, China}
   \affiliation{International Quantum Academy, Shenzhen, 518048, China}
   \author{Xiuqiang Wu}
   \affiliation{Department of Physics, Yancheng Institute of Technology, Yancheng 224051, China}
   \author{Yajie Ren}
   \affiliation{School of Physics and Telecommunication Engineering, Shaanxi University of Technology, Hanzhong 723001, China}
   \author{Jiansheng Wu}
   \email{wujs@sustech.edu.cn}
   \affiliation{Shenzhen Institute for Quantum Science and Engineering (SIQSE), Southern University of Science and Technology, Shenzhen 518055, China}
   \affiliation{International Quantum Academy, Shenzhen, 518048, China}

   \date{\today}

   \begin{abstract}
    Based on the Bogoliubov-de Gennes equations, we investigate the transport of the Josephson current in a S/$f_L$-F$_1$-$f_C$-F$_2$-$f_R$/S junction, where S and F$_{1,2}$ are superconductors and ferromagnets, and $f_{L, C, R}$ are the left, central, and right spin-active interfaces. These interfaces have noncollinear magnetizations, and the azimuthal angles of the magnetizations at the $f_{L, C, R}$ interfaces are $\chi_{L, C, R}$. We demonstrate that, if both the ferromagnets have antiparallel magnetizations, the critical current oscillates as a function of the exchange field and the thickness of the ferromagnets for particular $\chi_L$ or $\chi_R$. By contrast, when the magnetization at the $f_C$ interface is perpendicular to that at the $f_L$ and $f_R$ interfaces, the critical current reaches a larger value and is hardly affected by the exchange field and the thickness. Interestingly, if both the ferromagnets are converted to antiparallel half-metals, the critical current maintains a constant value and rarely changes with the ferromagnetic thicknesses and the azimuthal angles. At this time, an anomalous supercurrent can appear in the system, in which case the Josephson current still exists even if the superconducting phase difference $\phi$ is zero. This supercurrent satisfies the current-phase relation $I=I_c\sin(\phi+\phi_0)$ with $I_c$ being the critical current and $\phi_0=2\chi_C-\chi_L-\chi_R$. We deduce that the additional phase $\phi_0$ arises from phase superposition, where the phase is captured by the spin-triplet pairs when they pass through each spin-active interface. In addition, when both the ferromagnets are transformed into parallel half-metals, the $f_C$ interface never contributes any phase to the supercurrent and $\phi_0=\chi_R-\chi_L+\pi$. In such a case, the current-phase relation is similar to that in a S/$f_L$-F-$f_R$/S junction.
   \end{abstract}

    \maketitle

   \section{Introduction}

    In recent years great attention was paid to the study of superconductor (S)/ferromagnet (F) heterostructures due to plenty of fascinating phenomena that have been predicted and observed~\cite{Golubov,Buzdin,Bergeret,Linder,Eschg,Eschrig,JLAVBa}. It is well known that ferromagnetism and spin-singlet superconductivity are two inimical orders, as ferromagnetism favors a parallel spin alignment, while spin-singlet Cooper pairs consist of electrons with antiparallel aligned spins. Consequently, the ferromagnetic exchange field will make a dephasing effect on the electrons of the spin-singlet pairs~\cite{Buzdin,Bergeret}, when a F is adjacent to a conventional s-wave S. In hybrid S/F structures with homogeneous magnetization, the spin-singlet pairs ($\uparrow\downarrow-\downarrow\uparrow$) are destroyed by the exchange field of the F layer, so that they penetrate the F layer over a rather short scale. Meanwhile, the spin-triplet pairs ($\uparrow\downarrow+\downarrow\uparrow$) generate at the S/F interface and rapidly decay in the F layer. In contrast, the magnetic inhomogeneities could mediate equal-spin triplet pairs where both electrons are in the same spin band---either the majority band for spin-up triplet pairs ($\uparrow\uparrow$) or the minority band for spin-down triplet pairs ($\downarrow\downarrow$). Such triplet pairs are immune to the exchange field and can therefore penetrate the ferromagnet over a long distance from the S/F interface that the spin-singlet pairs could not reach~\cite{Buzdin,Bergeret,Linder,Eschg,Eschrig,JLAVBa}.

    In a uniform ferromagnetic Josephson junction (S/F/S), the wave function of the Cooper pairs penetrates ferromagnet on a short distance of the order $\xi_F=\hbar{v_f}/2h$ for ballistic systems and $\xi_F=\sqrt{\hbar{D}/h}$ for diffusive ones, where $v_f$ is the Fermi velocity, $h$ is ferromagnetic exchange field, and $D$ is the electronic diffusion constant~\cite{Buzdin}. Meanwhile, this penetration is accompanied by oscillations of the wave function in space. As a result, the critical current will reverse sign with changing temperature and thickness of the F layer (see~\cite{Buzdin} and references cited therein). Besides, in the Josephson junction with a nonuniform exchange field, the long-range supercurrent is apparent because the equal-spin triplet pairs ($\uparrow\uparrow$) [or ($\downarrow\downarrow$)] occur in the F layer~\cite{Bergeret,Linder,Eschg,Eschrig}. The penetration depth of these triplet pairs into a ferromagnet is much longer than $\xi_F$ and may be of the order of the Cooper pair penetration length into a normal metal, $\xi_N=\hbar{v_f}/2\pi{T}$ for ballistic systems and $\xi_N=\sqrt{\hbar{D}/2\pi{T}}$ for diffusive ones, where $T$ is the temperature~\cite{Buzdin}.

    On the other hand, in a traditional ferromagnetic Josephson junction, the ground state energy usually switches between specific superconducting phase differences $\phi=0$ and $\pi$, and the current-phase relation is sinusoidal $I(\phi)=I_c\sin\phi$, with $I_c$ being the critical current flowing through the junction~\cite{Golubov}. In the presence of breaking chiral~\cite{VKrive} and time-reversal symmetry~\cite{Golubov} in the Cooper pairs tunneling process, a spontaneous supercurrent at zero phase difference can arise, and the Josephson ground state can be characterized by a superconducting phase difference $\phi_0$. This supercurrent satisfies the current-phase relation $I(\phi)=I_c\sin(\phi+\phi_0)$~\cite{Golubov}. This $\phi_0$-junction could introduce excellent opportunities to quantum computer bits. It has been reported that the $\phi_0$-junction can be realized in Josephson junctions with ferromagnetic interlayers or under an externally applied Zeeman field in the presence of spin-orbit coupling~\cite{VKrive,AAReyn,ABuzdi,YTanak,AZazun,AGMals,ABrune,TYokoy,FSBerg,FKonsc,GCampa,DKuzma,MMinut,WMayer}, as well as with inhomogeneous ferromagnetic interlayers without an external spin-orbit coupling~\cite{VBYVNa,RGrein,JFLiu,IKulagina,Moor,AMoor,SMironov,IVBobk,DSRabinovich,MASila,SPal,HMeng}.

    It was theoretically predicted several years ago that the long-range triplet pairs with equal spins could be induced by the inhomogeneous magnetization configuration in the Josephson junctions~\cite{FSBAFVKBE,AKaRISh,MEJKJCC,HouzetBuzdin,YAYTa,YAYSYTAAG,MEsTLof,AVGal,MSKal,VolkovEfetov2010,LTrifu,MALinder2010,MengWuZheng}. The prediction of the long-range penetration of the spin-triplet pairs into a ferromagnet was observed in multiple experiments~\cite{RSKeizerSTB,ISHCho,JWARobinsonJDSW,DSprungmannKW,MSAFCzeschka,TSKhaire,MAKhasa,CaroKK,WillMMWP,JAGlick,DMNBanerjee,BMNiedzielski,VAguilar,DSManzano,NBanJWARob, JAGVAgui, MEgilmez}. It is very much worth noting that Houzet \emph{et al.}~\cite{HouzetBuzdin} proposed an alternative Josephson junction geometry of the form S/$f_L$-F-$f_R$/S in which the magnetization directions of the $f_L$ and $f_R$ layers are noncollinear to the central F layer. They suggested measuring the critical current in the Josephson junction to demonstrate the equal-spin triplet pairs. Soon after, the Birge group~\cite{TSKhaire,MAKhasa,CaroKK} observed a long-range supercurrent in Josephson junctions of the form S/$f_L$-Co-Ru-Co-$f_R$/S, where the central Co-Ru-Co trilayer was a synthetic antiferromagnet. The thin Ru layer induces antiparallel exchange coupling between the domains in the two Co layers, leaving nearly zero net magnetization in the junctions. To qualitatively explain the above experiments, the theoretical works of Volkov \emph{et al.}~\cite{VolkovEfetov2010} and Trifunovic \emph{et al.}~\cite{LTrifu} utilizing the quasiclassical theory studied the spatial distribution of spin-singlet and spin-triplet pair amplitudes and the long-range spin-triplet Josephson current in the S/$f_{L}$-F$_{1}$-F$_{2}$-$f_{R}$/S junction.

    There are three questions to be solved urgently: (i) The experiments found that, without the Ru layer, the critical currents were very small~\cite{TSKhaire,MAKhasa,CaroKK}. However, in both theoretical works~\cite{VolkovEfetov2010,LTrifu}, the Ru layer was regarded as a non-magnetic metal and thus its influence on the transport of the spin-triplet pairs was not considered. (ii) It is known that the majority spin-triplet component has a larger amplitude than the minority spin-triplet component in the strong ferromagnet. In the S/$f_{L}$-F$_{1}$-F$_{2}$-$f_{R}$/S junction, when magnetizations of the F$_{1}$ and F$_{2}$ layers are antiparallel to each other, the majority spin-triplet pairs in the F$_{1}$ layer become minority spin-triplet pairs in the F$_{2}$ layer and vice versa. As a result, both triplet components suffer from the lower transmission amplitude of the minority component somewhere in the system. If that were the whole story, then the Josephson current should be small. In particular, the current is completely inhibited when the F$_{1}$ and F$_{2}$ layers are turned into antiparallel half-metals. This inference is inconsistent with the critical current enhancement observed in the experiments~\cite{TSKhaire,MAKhasa,CaroKK}. (iii) The quasiclassical approximation used by both theoretical works~\cite{VolkovEfetov2010,LTrifu} does not consider the fact that the transport properties of the majority and minority electrons at the Fermi surface in the F$_{1}$ and F$_{2}$ layers are very different, in which case the spontaneous supercurrent cannot be captured.

    The purpose of this paper is to address the three questions raised above. Recently,  Quarterman \emph{et al.}~\cite{PQuart} experimentally demonstrated that ferromagnetism can occur in ultrathin Ru films. This evidence can be used to address questions (i) and (ii) mentioned above. The Ru layer with noncollinear magnetizations can induce a spin-flip scattering effect, which converts the spin-up (spin-down) triplet pairs in the F$_{1}$ layer to the spin-down (spin-up) triplet pairs in the F$_{2}$ layer. In this process, the Ru layer acts as a bridge connecting the F$_{1}$ and F$_{2}$ layers to ensure supercurrent transmission. To address question (iii), we employ a microscopical approach to solve the Bogoliubov--de Gennes (BdG) equations~\cite{PGdeGennes}. The exact numerical solutions of these equations can acquire the spontaneous supercurrent.

    In this paper, we study the propagation of long-range Josephson current in the S/$f_L$-F$_1$-$f_C$-F$_2$-$f_R$/S junction, where F$_{1, 2}$ are antiparallel or parallel ferromagnets, and $f_{L, C, R}$ denote the left, central, and right spin-active interfaces. All these interfaces have noncollinear magnetizations, and the azimuthal angles of the magnetizations at the $f_{L, C ,R}$ interfaces are $\chi_{L, C, R}$. When the F$_1$ and F$_2$ layers are antiparallel to each other, the Josephson critical current exhibits different characteristics for the different azimuthal angles of the interfaces. If the azimuthal angle $\chi_L$ or $\chi_R$ takes particular values, the critical current will oscillate with increasing exchange field and thickness of the F$_1$ and F$_2$ layers. For other values of $\chi_L$ or $\chi_R$, the oscillation effect of the current hardly appears. By comparison, when the magnetization of the $f_C$ interface is perpendicular to that of the $f_L$ and $f_R$ interfaces, the critical current reaches a higher value and is essentially unaffected by the exchange field and thickness.

    Interestingly, when both ferromagnetic layers turn into antiparallel half-metals, in which case the very large magnetization strength permits only one spin to exist, the critical current always remains a constant value and is rarely affected by the ferromagnetic thicknesses and the azimuth angles. At this time, the Josephson current gains an additional phase $\phi_0=2\chi_C-\chi_L-\chi_R$ to form the spontaneous supercurrent. We consider that the phase $\phi_0$ arises from the phase superposition effect, where the phase is obtained by the spin-triplet pairs when they transport through each spin-active interface. In contrast, if the F$_1$ and F$_2$ layers become parallel half-metals, the central $f_C$ interface is similar to an ordinary spin-independent barrier and never makes any contribution to the phase $\phi_0$. It can be attributed to the phase cancellation of the spin-triplet pairs when they pass through the $f_C$ interface. As a result, the current-phase relation is similar to that in the S/$f_L$-F-$f_R$/S junction. The advantages of the S/$f_L$-F$_1$-$f_C$-F$_2$-$f_R$/S junction and the possible experimental implementation are introduced in the Supplemental Material~\cite{SupMate}.

   \section{Model and formula} \label{Sec2}

   In general, Green's function technique is a very powerful tool for studying diffusive S/F systems. The quasiclassical approaches for Green's function were proposed by Eilenberger~\cite{GEilen} and Usadel~\cite{KDUsadel} successively. However, the applicability of these methods assumes that the exchange field $h$ in the ferromagnet should be much smaller than the Fermi energy $h\ll{E_{F}}$ and the use of the Usadel equations implies even more restrictive conditions $h\tau\ll1$, where $\tau$ is the electrons scattering time~\cite{Buzdin}. As a result, since the transport properties of the majority and minority electrons in the F layers are very different, the quasiclassical approaches lose their effectiveness, and some subtle qualitative effects may be missed, see, for example, \cite{CRReeg,MASila,HMeng,HMBuzdin}. Moreover, a lot of experimental activities with the S/F heterostructures deal with strong ferromagnets (or even half-metals~\cite{CVisani2012,CVisani2015,MEgilmez}) for which the quasiclassical approximation cannot provide an adequate quantitative description. The alternative approach for the description of proximity effects in strong ferromagnets is the use of the microscopical approach based on the BdG equations~\cite{PGdeGennes}. For the inhomogeneous ferromagnetic Josephson junction, analytical solutions to the BdG equations are generally not easy to obtain. The exact numerical solutions of these equations may provide additional information to the quasiclassical approach and this method was used in~\cite{ZRadovi,KHalt2004,ZPajovi,PHBar2007,KHPHBOTV2007,KHOTV2009,KHOTVCTW,KHalterman2016,KHMAlidoust,CTWuKHa,MAliKHal2018,KHalMAl2018,KHMARSmith} and references cited therein. In the following, we describe the generalized BdG method in detail.

\begin{figure}[ptb]
       \centering
       \includegraphics[width=3.2in]{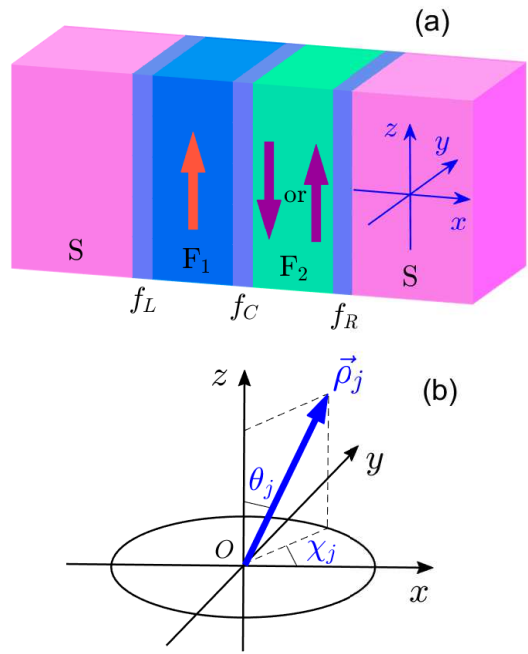}
       \caption{(a) Schematic diagram of the S/$f_L$-F$_{1}$-$f_C$-F$_{2}$-$f_R$/S junction, where thick arrows in F$_1$ and F$_2$ indicate the directions of the exchange field, and the thicknesses of F$_1$ and F$_2$ are $d_1$ and $d_2$, respectively. The $f_{L}$, $f_{C}$, and $f_R$ interfaces are assumed to be spin-active potential barriers due to misaligned local magnetizations. (b) The direction of magnetization $\vec{\rho}_{j}$ at the $f_{j}$ interface, where $j=L$, $C$, and $R$ correspond to the left, central, and right interfaces, respectively. $\theta_{j}$ and $\chi_{j}$ denote polar and azimuthal angles.}
       \label{Fig1}
    \end{figure}

   The considered S/$f_L$-F$_{1}$-$f_C$-F$_{2}$-$f_R$/S Josephson junction is shown schematically in Fig.~\ref{Fig1}. The $x$ axis is chosen to be perpendicular to the layer interfaces with the origin located at the position of the $f_{C}$ interface. The BCS mean-field effective Hamiltonian is given by~\cite{Buzdin,PGdeGennes}
   \begin{align}
     H_{\rm {eff}}=&{\displaystyle\sum\limits_{\alpha,\beta}}\int{d}\mathbf{r}\left\{
     \hat{\psi}_{\alpha}^{\dagger}(\mathbf{\mathbf{r}})\left[H_{e}-(h_{z}
     \hat{\sigma}_{z})_{\alpha\alpha}\right]\hat{\psi}_{\alpha}(\mathbf{r})\right.
     \nonumber\\
     &+\frac{1}{2}\left[(i\hat{\sigma}_{y})_{\alpha\beta}\Delta(\mathbf{r}
     )\hat{\psi}_{\alpha}^{\dagger}(\mathbf{r})\hat{\psi}_{\beta}^{\dagger}(\mathbf{r})+{\rm H.c.}\right] \nonumber \\
     &-\left.\hat{\psi}_{\alpha}^{\dagger}(\mathbf{r})\left(\vec{\rho}\cdot\vec{\sigma}\right)_{\alpha\beta}\hat{\psi}_{\beta}(\mathbf{r})\right\} ,\label{HEFF}
   \end{align}
   where $H_{e}=-\frac{\hbar^{2}\nabla^{2}}{2m}-E_{F}$, and $\hat{\psi}_{\alpha}^{\dagger}(\mathbf{r})$ and $\hat{\psi}_{\alpha}(\mathbf{r})$ represent creation and annihilation operators with spin $\alpha$. Here $\vec{\sigma}=\left(\hat{\sigma}_{x},\hat{\sigma}_{y},\hat{\sigma}_{z}\right)$ is the vector of Pauli matrices, $m$ denotes the effective mass of the quasiparticles in both the superconductors and the ferromagnets, and $E_{F}$ is the Fermi energy. We assume equal Fermi energies in the different regions of the junction. The superconducting gap is supposed to be constant in the superconducting electrodes and absent inside the ferromagnetic region:
   \begin{equation}
      \Delta(\mathbf{r})=
         \begin{cases}
            {\Delta}e^{i\phi/2}, & x<-d_{1},\\
            0, & -d_{1}<x<d_{2}, \\
            {\Delta}e^{-i\phi/2}, & x>d_{2},
         \end{cases}
    \end{equation}
     where $\Delta$ is the magnitude of the gap, and $\phi$ is the phase difference between the two superconducting electrodes. This approximation is justified when, for example, the thickness of the superconducting layers is much larger than the thickness of ferromagnetic layers. The exchange field in two ferromagnetic layers is parallel or antiparallel to the $z$ axis. It has the form
     \begin{equation}
        {h}_{z}=
        \begin{cases}
           h_{1}\hat{z},  & -d_{1}<x<0, \\
           \pm{h_{2}\hat{z}}, & 0<x<d_{2},
        \end{cases}
     \end{equation}
     where $\hat{z}$ is the unit vector along the $\emph{z}$ axis.
     We model the spin-active interface by a $\delta$ function potential barrier $\vec{\rho}(x)=\vec{\rho}_{L}\delta(x+d_{1})+\vec{\rho}_{C}\delta(x)+\vec{\rho}_{R}\delta(x-d_{2})$, where $\vec{\rho}_{j}$ ($j=L$, $C$ and $R$) is a vector parallel to the magnetization at the $f_{j}$ interface. The components of $\vec{\rho}_{j}$ are characterized by the polar angle $\theta_j$ and the azimuthal angle $\chi_j$ in the usual way
     \begin{equation}
        \begin{cases}
           \rho_{j}^{x}=\rho_{j}\sin\theta_{j}\cos\chi_{j}, \\
           \rho_{j}^{y}=\rho_{j}\sin\theta_{j}\sin\chi_{j}, \\
           \rho_{j}^{z}=\rho_{j}\cos\theta_{j},
        \end{cases}
     \end{equation}
     where $\rho_{j}$ represents the strength of the interfacial magnetization.

    To diagonalize the effective Hamiltonian $H_{\rm {eff}}$, the field operator $\hat{\psi}_\alpha(\mathbf{r})$ is expanded utilizing the Bogoliubov transformation $\hat{\psi}_{\alpha}(\mathbf{r})=\sum_{n}[u_{n\alpha}(\mathbf{r})\hat{\gamma}_{n}+v_{n\alpha}^{\ast}(\mathbf{r})\hat{\gamma}_{n}^{\dag}]$, where $u_{n\alpha}$ and $v_{n\alpha}$ represent the quasiparticle amplitude, and $\hat{\gamma}_{n}$, $\hat{\gamma}_{n}^{\dag}$ are the Bogoliubov annihilation and creation operators, respectively. Using the presentation $u_{n\alpha}(\mathbf{r})=u_{k}^{\alpha}e^{ikx}$, $v_{n\alpha}(\mathbf{r})=v_{k}^{\alpha}e^{ikx}$, the resulting BdG equations can be expressed as~\cite{PGdeGennes}
    \begin{equation}%
        \begin{pmatrix}
           \hat{H}_{0}+\hat{U} & i\hat{\sigma}_{y}\Delta(x)\\
           -i\hat{\sigma}_{y}\Delta^{\ast}(x) & -\hat{H}_{0}-\hat{U}^{\ast}
        \end{pmatrix}
        \begin{pmatrix}
           \hat{u}(x)\\
           \hat{v}(x)
        \end{pmatrix}
        =\epsilon
        \begin{pmatrix}
           \hat{u}(x)\\
           \hat{v}(x)
        \end{pmatrix},\label{BdG}
    \end{equation}
    where
    \[
    \hat{H}_{0}=
    \begin{pmatrix}
        \xi_{k}-{h}_{z} & 0\\
        0 & \xi_{k}+{h}_{z}
    \end{pmatrix},
    \]
    \begin{equation}
       \hat{U}=\hat{U}_{L}\delta(x+d_{1})+\hat{U}_{C}\delta(x)+\hat{U}_{R}\delta(x-d_{2}), \nonumber
    \end{equation}
    and
    \[
    \hat{U}_{j}=
    \begin{pmatrix}
        -{\rho}_{j}^{z} & -(\rho_{j}^{x}-i\rho_{j}^{y})\\
        -(\rho_{j}^{x}+i\rho_{j}^{y}) & {\rho}_{j}^{z}
    \end{pmatrix}.
    \]
    Here $\xi_{k}=\frac{\hbar^{2}k^{2}}{2m}-E_{F}$, and $\hat{u}(x)=[u_{k}^{\uparrow}(x)\;u_{k}^{\downarrow}(x)]^{T}$ and $\hat{v}(x)=[v_{k}^{\uparrow}(x)\;v_{k}^{\downarrow}(x)]^{T}$ are quasiparticle and quasihole wave functions, respectively.

    The BdG equations (\ref{BdG}) can be solved for each superconducting electrode and each ferromagnetic layer, respectively. For a given energy $\epsilon$ in the superconducting gap, we find the following plane-wave solutions in the left superconducting electrode:
    \begin{align}
       \psi_{L}^{S}(x)&=C_{1}\hat{\zeta}_{1}e^{-ik_{S}^{+}x}+C_{2}\hat{\zeta}_{2}e^{ik_{S}^{-}x} \nonumber \\
       &+C_{3}\hat{\zeta}_{3}e^{-ik_{S}^{+}x}+C_{4}\hat{\zeta}_{4}e^{ik_{S}^{-}x}, \label{functionSL}
    \end{align}
    where $k_{S}^{\pm}=k_{F}\sqrt{1\pm{i}\sqrt{\Delta^{2}-\epsilon^{2}}/E_{F}-(k_{\parallel}/k_{F})^{2}}$ are the perpendicular components of the wave vectors for quasiparticles in both superconductors, and $k_{\|}$ is the parallel component. $\hat{\zeta}_{1}=[1\;0\;0\;R_{1}e^{-i\phi/2}]^{T}$, $\hat{\zeta}_{2}=[1\;0\;0\;R_{2}e^{-i\phi/2}]^{T}$, $\hat{\zeta}_{3}=[0\;1\;-R_{1}e^{-i\phi/2}\;0]^{T}$, and $\hat{\zeta}_{4}=[0\;1\;-R_{2}e^{-i\phi/2}\;0]^{T}$ are the four basis wave functions of the left superconductor, in which $R_{1(2)}=(\epsilon\mp{i}\sqrt{\Delta^{2}-\epsilon^{2}})/\Delta$. The corresponding wave function in the right superconducting electrode can be described by
    \begin{align}
       \psi_{R}^{S}(x)&=D_{1}\hat{\eta}_{1}e^{ik_{S}^{+}x}+D_{2}\hat{\eta}_{2}e^{-ik_{S}^{-}x} \nonumber \\
       &+D_{3}\hat{\eta}_{3}e^{ik_{S}^{+}x}+D_{4}\hat{\eta}_{4}e^{-ik_{S}^{-}x}, \label{functionSR}
    \end{align}
    where $\hat{\eta}_{1}=[1\;0\;0\;R_{1}e^{i\phi/2}]^{T}$, $\hat{\eta}_{2}=[1\;0\;0\;R_{2}e^{i\phi/2}]^{T}$, $\hat{\eta}_{3}=[0\;1\;-R_{1}e^{i\phi/2}\;0]^{T}$, and $\hat{\eta}_{4}=[0\;1\;-R_{2}e^{i\phi/2}\;0]^{T}$.

    The wave function in the F$_{1}$ layer is
    \begin{align}
       \psi_{1}(x)&=(M_{1}e^{ik_{1}x}+M_{1}^{\prime}e^{-ik_{1}x})\hat{e}_{1} \nonumber\\
       &+(M_{2}e^{ik_{2}x}+M_{2}^{\prime}e^{-ik_{2}x})\hat{e}_{2}  \nonumber\\
       &+(M_{3}e^{ik_{3}x}+M_{3}^{\prime}e^{-ik_{3}x})\hat{e}_{3}  \nonumber\\
       &+(M_{4}e^{ik_{4}x}+M_{4}^{\prime}e^{-ik_{4}x})\hat{e}_{4},\label{HM_wave}
      \end{align}
    where $\hat{e}_{1}=(1\;0\;0\;0)^{T}$, $\hat{e}_{2}=(0\;1\;0\;0)^{T}$, $\hat{e}_{3}=(0\;0\;1\;0)^{T}$, and $\hat{e}_{4}=(0\;0\;0\;1)^{T}$ are basis wave functions in the ferromagnetic region, and $k_{1(2)}=k_{F}\sqrt{1+(\epsilon \pm{h_{1}})/E_{F}-\left( k_{\parallel}/k_{F}\right) ^{2}}$ and $k_{3(4)}=k_{F}\sqrt{1-(\epsilon\mp{h_{1}})/{E_{F}}-\left( k_{\parallel}/{k_{F}}\right)^{2}}$ are the perpendicular components of the wave vectors for the quasiparticles in the F$_{1}$ layer. The corresponding wave function $\psi_{2}(x)$ in the F$_{2}$ layer can be derived from Eq.~(\ref{HM_wave}) by replacement $h_{1}\rightarrow{h_{2}}$. It is worthy to note that the parallel component $k_{\parallel}$ is conserved in the transport processes of the quasiparticles.

    \begin{figure*}
        \centering
        \includegraphics[width=7.0in]{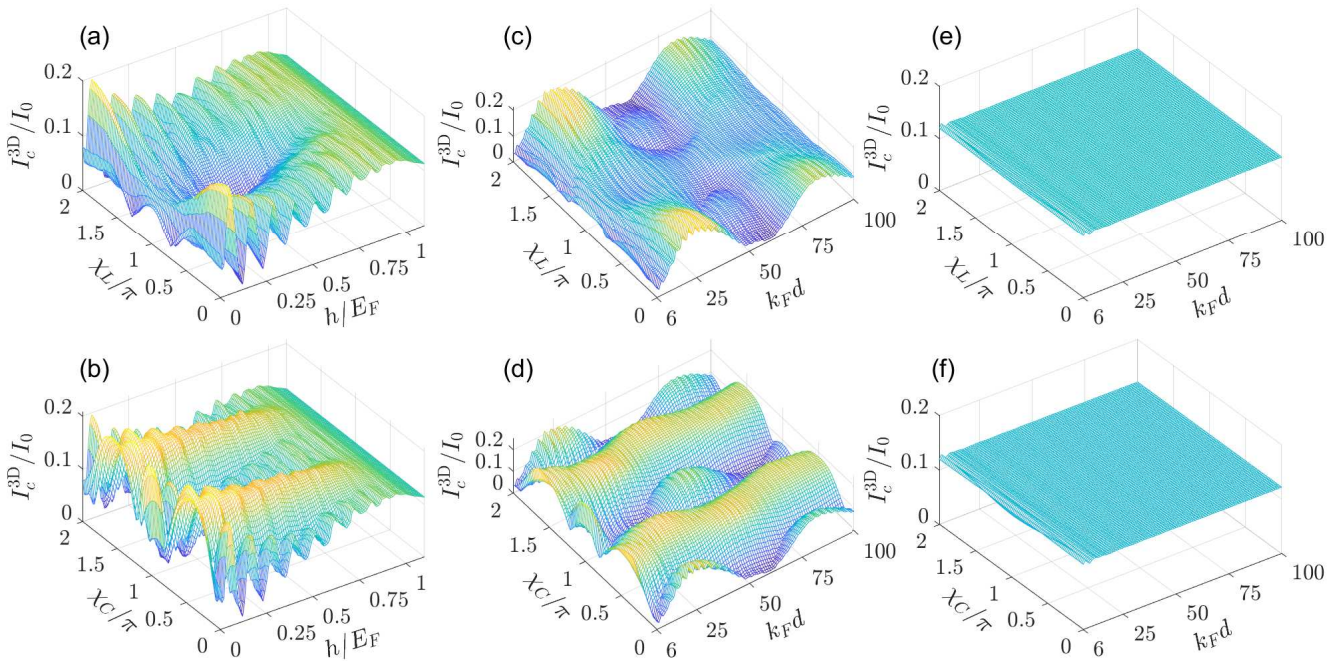} 
        \caption{The critical current $I_{c}^{3\rm{D}}$ versus the left azimuthal angle $\chi_{L}$ for $\chi_{C}=\chi_{R}=0$ [(a), (c), and (e)], and $I_{c}^{3\rm{D}}$ versus $\chi_{C}$ for $\chi_{L}=\chi_{R}=0$ [(b), (d), and (f)]. Here it is shown that $I_{c}^{3\rm{D}}$ varies with the exchange field $h/E_F$ for the ferromagnetic thickness $k_Fd=50$ [(a) and (b)], and $I_{c}^{3\rm{D}}$ varies with $k_Fd$ for $h/E_F=0.1$ [(c) and (d)] and $h/E_F=1.05$ [(e) and (f)]. In all panels, the spin-active barriers are taken as $P_{L}=P_{R}=1$ and $P_{C}=1.5$.}
        \label{Fig2}
    \end{figure*}

    The wave functions $\big[$$\psi_{L}^{S}(x)$, $\psi_{1}(x)$, $\psi_{2}(x)$, and $\psi_{R}^{S}(x)$$\big]$ and their first derivatives satisfy the following boundary conditions
    \begin{align}
         & \psi_{L}^{S}(-d_{1})=\psi_{1}(-d_{1}),\nonumber\\
         & \psi^{\prime}_{1}(-d_{1})-\psi_{L}^{\prime S}(-d_{1})=k_{F}
         \begin{pmatrix}
            \hat{V}_{L} & 0\\
            0 & \hat{V}_{L}^{*}
         \end{pmatrix}
         \psi(-d_1), \label{boundary1}\\
         & \psi_{1}(0)=\psi_{2}(0),\nonumber\\
         & \psi^{\prime}_{2}(0)-\psi^{\prime}_{1}(0)=k_{F}
         \begin{pmatrix}
            \hat{V}_{C} & 0\\
            0 & \hat{V}_{C}^{*}
         \end{pmatrix}
         \psi(0), \label{boundary2}\\
         & \psi_{2}(d_2)=\psi_{R}^{S}(d_{2}), \nonumber\\
         & \psi_{R}^{\prime S}(d_{2})-\psi^{\prime}_{2}(d_{2})=k_{F}
         \begin{pmatrix}
            \hat{V}_{R} & 0\\
            0 & \hat{V}_{R}^{*}
         \end{pmatrix}
         \psi(d_2), \label{condition3}
    \end{align}
     where
    \begin{equation}
        \hat{V}_{j}=
        \begin{pmatrix}
            -P_{j}^{z} & -(P_{j}^{x}-iP_{j}^{y})\\
            -(P_{j}^{x}+iP_{j}^{y}) & P_{j}^{z}
        \end{pmatrix}.
    \end{equation}
    We define the dimensionless spin-dependent parameter $(P_{j}^{x},P_{j}^{y},P_{j}^{z})=P_{j}(\sin\theta_{j}\cos\chi_{j},\sin\theta_{j}\sin\chi_{j},\cos\theta_{j})$, where the dimensionless parameter $P_{j}=2m\rho_{j}/(\hbar^{2}k_{F})$ describes the strength of the spin-active barrier at the $f_{j}$ interface.

    From these boundary conditions, we can set up 24 linear equations in the following form:
    \begin{equation}
       \hat{A}X=\hat{B},\label{linearEq}
    \end{equation}
    where $X$ contains 24 scattering coefficients, and $\hat{A}$ is a $24\times24$ matrix. The solution of the characteristic equation
    \begin{equation}
       \det\hat{A}=0\label{characteristicEq}
    \end{equation}
    allows one to identify two Andreev bound-state solutions for energies $E_{A\omega}$ ($\omega$=1, 2). Below we will consider the case of the short Josephson junction with a thickness much smaller than the superconducting coherence length $\xi_S$. In such a case, the contribution to the Josephson current comes from the discrete Andreev bound states, and the continuous electron state does not play any role (see, e.g.,~\cite{PFBagwell, CWJBeenakker}). In a one-dimensional (1D) structure, the Josephson current can be calculated by the general formula
    \begin{equation}
       I^{1\rm{D}}(\phi)=\frac{2e}{\hbar}\frac{\partial\Omega}{\partial\phi},\label{current}
    \end{equation}
    where $\Omega$ is the phase-dependent thermodynamic potential. This potential arises from the excitation spectrum by using the formula~\cite{JBardeen,JCayssol}
    \begin{equation}
        \Omega=-2T\sum_{\omega}\ln\left[ 2\cosh\frac{E_{A\omega}(\phi)}{2T}\right],\label{potential}
    \end{equation}
    where $\Delta$, $h_{1}$, $h_{2}$, $P_{j}$, $\theta_{j}$, and $\chi_{j}$ are assumed to be the equilibrium values, which minimize the free energy of the S/$f_L$-F$_{1}$-$f_C$-F$_{2}$-$f_R$/S junction and depend on microscopic parameters~\cite{Buzdin-AdvPhys85}. The summation in (\ref{potential}) is taken over all positive Andreev energies \big[$0<E_{A\omega}(\phi)<\Delta$\big]. For each value of $\phi$, we solve Eq.~(\ref{characteristicEq}) numerically to obtain the two spin-polarized Andreev levels. Since the Andreev energy spectra are doubled as they include the Bogoliubov redundancy, and only half of the energy states should be taken into account, we can find the 1D Josephson current via Eqs.~(\ref{current}) and (\ref{potential}).

    In a three-dimensional (3D) case, the Josephson current can be expressed as~\cite{HMBuzdin}
    \begin{equation}
        I^{3\rm{D}}(\phi)=\frac{4\pi\Delta}{eR_{N}}\int^{1}_{0}{I}^{\rm{1D}}(\tilde{k}_{\parallel}){\tilde{k}_{\parallel}}d\tilde{k}_{\parallel},\label{3Dcurrent}
    \end{equation}
    where $R^{-1}_{N}=e^{2}k^{2}_{F}S/(4\pi^{2}\hbar)$ is the Sharvin resistance, and $\tilde{k}_{\parallel}=k_{\parallel}/k_{F}$ is the normalized wave vector. The 3D critical current can be derived from $I_{c}^{3\rm{D}}=\max_{\phi}|I^{3\rm{D}}(\phi)|$.

    \section{Results and discussions}  \label{Sec3}

      In our calculations we use the superconducting gap $\Delta$ as a unit of energy and take the Fermi energy $E_{F}=1000\Delta$. All length scales and the exchange field strengths are normalized by the inverse Fermi wave-vector $k_{F}$ and the Fermi energy $E_{F}$, respectively. Note that the approximation of the short Josephson junction ($k_Fd_1,k_Fd_2\ll1000$) is fully satisfied in the presented calculations. The normalized unit of current is $I_{0}=2\pi\Delta/(eR_{N})$ in the 3D case. The calculations of all the currents are performed at temperature $T=0$ throughout the paper.

      \begin{figure*}[ptb]
        \centering
        \includegraphics[width=7.0in]{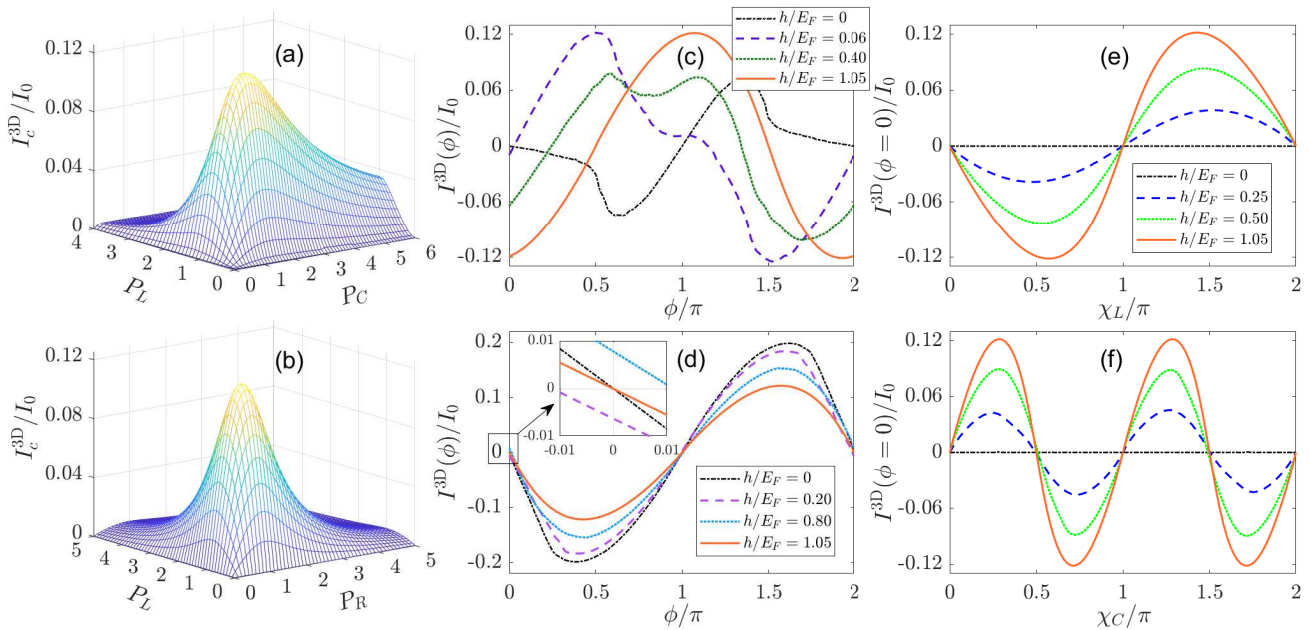}
        \caption{The critical current $I_{c}^{3\rm{D}}$ versus the strength of the spin-active barriers $P_{L}$ and $P_{C}$ in the case of $P_{L}=P_{R}$ (a), and $I_{c}^{3\rm{D}}$ versus $P_{L}$ and $P_{R}$ for $P_{C}=1.5$ (b). In (a) and (b) we choose $h/E_{F}=1.05$, $k_Fd=50$, $\chi_{L}=\chi_{R}=0$, and $\chi_C=\pi/2$. Current-phase relation $I^{\rm{3D}}(\phi)$ under different exchange fields $h/E_F$ for $\chi_L=\pi/2$ when $\chi_C=\chi_R=0$ (c) and for $\chi_C=\pi/2$ when $\chi_L=\chi_R=0$ (d). Spontaneous supercurrent $I^{\rm{3D}}(\phi=0)$ versus $\chi_L$ for $\chi_C=\chi_R=0$ (e), and $I^{\rm{3D}}(\phi=0)$ versus $\chi_C$ for $\chi_L=\chi_R=0$ (f) when $h/E_F$ takes several values. (e) and (f) use the same legends.  In (c)--(f) we choose $k_Fd=50$, $P_L=P_R=1$, and $P_C=1.5$.}
        \label{Fig3}
      \end{figure*}

      For simplicity, we define that the magnetization vector $\vec{\rho}_j$ at the $f_j$ interface lies in the $x$-$y$ plane ($\theta_L=\theta_C=\theta_R=\pi/2$), and sets up the azimuthal angle $\chi_j$ with the $x$ axis, see Fig.~\ref{Fig1}. The magnetization of the F$_1$ layer is fixed along the $z$ axis, and the direction of the F$_2$ layer is antiparallel or parallel to the F$_1$ layer. Unless otherwise stated, the results are calculated for the same strength of exchange fields ($h_{1}=\pm{h_{2}}=h$) and the same thicknesses ($d_{1}=d_{2}=d$) in the F$_1$ and F$_2$ layers.

      \subsection{The Josephson current in the S/$f_L$-F$_{1}$-$f_C$-F$_{2}$-$f_R$/S junction with antiparallel magnetization configurations}

      We discuss first the Josephson current for the antiparallel magnetic configurations ($h_{1}=-h_{2}=h$). As illustrated in Figs.~\ref{Fig2}(a) and \ref{Fig2}(c), $I_{c}^{3\rm{D}}$ exhibits an oscillatory characteristic with increasing $h/E_F$ and $k_Fd$. The oscillating effect of $I_{c}^{3\rm{D}}$ is more prominent at $\chi_L=0$ and $2\pi$ but almost no longer appears at $\chi_L=\pi$. Figures~\ref{Fig2}(b) and \ref{Fig2}(d) show that when $\chi_{C}$ takes values around $0.5\pi$ and $1.5\pi$, $I_{c}^{3\rm{D}}$ always maintains a large amplitude over the entire range of $h/E_F$ and $k_Fd$. As $\chi_{C}$ is in the vicinity of $0$, $\pi$, and $2\pi$, the $I_{c}^{3\rm{D}}$ amplitude decreases, but its oscillatory character with $h/E_F$ and $k_Fd$ becomes more pronounced. For the half-metallic ferromagnet in Figs.~\ref{Fig2}(e) and \ref{Fig2}(f), $I_{c}^{3\rm{D}}$ neither changes with the azimuthal angles $\chi_{L}$ and $\chi_{C}$ nor with the ferromagnetic thickness $k_Fd$. These features demonstrate that the Josephson current through the system is a long-range spin-polarized supercurrent. Moreover, the left and right interfaces play the same role in the transport process of the Josephson current. The variation of the Josephson current with the two interface angles ($\chi_L$ and $\chi_R$) is described in the Supplemental Material~\cite{SupMate}.

      Continually, we explore the dependence of the Josephson current on the interfacial barriers and the ferromagnetic exchange fields. As shown in Fig.~\ref{Fig3}(a), without the spin-active barriers ($P_L=0$ or/and $P_C=0$), $I^{\rm{3D}}_c$ is zero. As the interfacial barriers increase, $I^{\rm{3D}}_c$ increases rapidly and reaches a maximum for $P_L=1$ and $P_C=1.5$, then decreases at larger values of $P_L$ and $P_C$. The decrease in $I^{\rm{3D}}_c$ at large $P_L$ and $P_C$ signals that the interfacial barriers not only flip the electron spin but also suppress the transport of paired electrons. By contrast, as illustrated in Fig.~\ref{Fig3}(b), $I^{\rm{3D}}_c$ shows the same dependence on the left and right barriers and reaches a maximum at $P_L=P_R=1$. On the other hand, Fig.~\ref{Fig3}(c) shows the dependence of $I^{\rm{3D}}(\phi)$ on $h/E_F$, when the $f_L$ interface is magnetized along the $y$ axis, and the other two interfaces are along the $x$ axis. Under a weak $h/E_F$, $I^{\rm{3D}}(\phi)$ cannot be represented as a sinusoidal function. But when $h/E$ is strong enough, the current satisfies a relation $I^{\rm{3D}}(\phi)=I^{\rm{3D}}_c\sin(\phi-\pi/2)$ for $\chi_L=\pi/2$. When the $f_C$ interface is oriented along the $y$ axis, the $I^{\rm{3D}}(\phi)$ amplitude gradually decreases with increasing $h/E_F$, but the current approximately maintains a sinusoidal relation $I^{\rm{3D}}(\phi)=I^{\rm{3D}}_c\sin(\phi+\pi)$ for $\chi_C=\pi/2$ [see Fig.~\ref{Fig3}(d)]. In comparison, as shown in Figs.~\ref{Fig3}(e) and \ref{Fig3}(f), $I^{\rm{3D}}(\phi=0)$ does not exist at $h/E_F=0$, and its amplitude increases with increasing $h/E_F$. As the two ferromagnets are converted into half-metals, the amplitude reaches the maximum. The oscillation periods of $I^{\rm{3D}}(\phi=0)$ with angles $\chi_L$ and $\chi_C$ are $2\pi$ and $\pi$, respectively. Therefore, the spontaneous supercurrent satisfies the following relations $I^{\rm{3D}}(\phi=0)=I^{\rm{3D}}_c\sin(-\chi_L)$ for $\chi_C=\chi_R=0$ and $I^{\rm{3D}}(\phi=0)=I^{\rm{3D}}_c\sin(2\chi_C)$ for $\chi_L=\chi_R=0$.

      \begin{figure*}
         \begin{minipage}{0.72\textwidth}
           \includegraphics[width=0.95\textwidth]{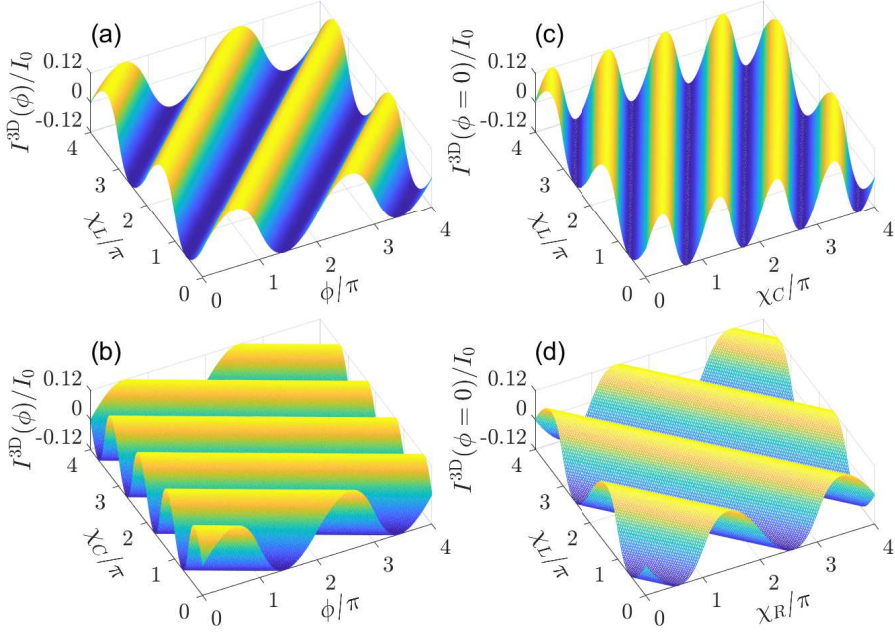}
         \end{minipage}%
         \begin{minipage}{0.28\textwidth}
            \caption{Current-phase relation $I^{\rm{3D}}(\phi)$ versus the azimuthal angles $\chi_L$ (a) and $\chi_C$ (b). Spontaneous supercurrent $I^{\rm{3D}}(\phi=0)$ versus the azimuthal angles ($\chi_L$, $\chi_C$) (c) and ($\chi_L$, $\chi_R$) (d). Here the unlabeled azimuthal angles in each panel are set to 0, and the parameters are as follows: $h/E_F=1.05$, $k_Fd=50$, $P_L=P_R=1$, and $P_C=1.5$.}
            \label{Fig4}
         \end{minipage}
      \end{figure*}

       Next, we show the current-phase relation $I^{\rm{3D}}(\phi)$ and the spontaneous supercurrent $I^{\rm{3D}}(\phi=0)$ for the half-metallic phase in Fig.~\ref{Fig4}. It can be seen that the oscillation period of current with $\chi_L$ and $\chi_R$ is twice that with $\chi_C$. According to the inference described in the Supplemental Material~\cite{SupMate}, we can deduce a complete current-phase relation $I^{\rm{3D}}(\phi)=I^{\rm{3D}}_c\sin(\phi+2\chi_C-\chi_L-\chi_R)$ for the entire system.

       We now give a simple physical picture to describe the propagation process of the Cooper pairs in the S/$f_L$-F$_{1}$-$f_C$-F$_{2}$-$f_R$/S junction when the F$_1$ and F$_2$ layers are antiparallel half-metals. As mentioned before, the orientation of the interfacial barrier is characterized by the polar ($\theta$) and the azimuthal ($\chi$) angles measured from the $z$ axis in spin space. The transformation formulas for basis vectors quantized along the direction ($\theta$, $\chi$) in terms of basis vectors quantized along the $z$ axis read~\cite{Eschrig}
        \begin{subequations}
         \begin{align}
            (\uparrow)_{\theta,\chi}&=\cos\frac{\theta}{2}e^{-i\frac{\chi}{2}}(\uparrow)_{z}+\sin\frac{\theta}{2}e^{i\frac{\chi}{2}}(\downarrow)_{z}, \label{Za}\\
            (\downarrow)_{\theta,\chi}&=-\sin\frac{\theta}{2}e^{-i\frac{\chi}{2}}(\uparrow)_{z}+\cos\frac{\theta}{2}e^{i\frac{\chi}{2}}(\downarrow)_{z}, \label{Zb}
         \end{align}
        \end{subequations}
        and can be used to find the transformation for the spin-singlet pairs and the spin-triplet pairs~\cite{Eschrig}
        \begin{subequations}
            \begin{align}
            (\uparrow\downarrow-\downarrow\uparrow)_{\theta,\chi}=&(\uparrow\downarrow-\downarrow\uparrow)_{z}, \label{FLa}\\
            (\uparrow\downarrow+\downarrow\uparrow)_{\theta,\chi}=&-\sin\theta\left[e^{-i\chi}(\uparrow\uparrow)_{z}-e^{i\chi}(\downarrow\downarrow)_{z}\right]  \nonumber \\
                                                                  &+\cos\theta(\uparrow\downarrow+\downarrow\uparrow)_{z}. \label{FLb}
             \end{align}
        \end{subequations}
        The reverse transformation formulas can be written as
        \begin{subequations}
             \begin{align}
               (\uparrow)_{z}=&\cos\frac{\theta}{2}e^{i\frac{\chi}{2}}(\uparrow)_{\theta,\chi}-\sin\frac{\theta}{2}e^{i\frac{\chi}{2}}(\downarrow)_{\theta,\chi}, \label{Za}\\
               (\downarrow)_{z}=&\sin\frac{\theta}{2}e^{-i\frac{\chi}{2}}(\uparrow)_{\theta,\chi}+\cos\frac{\theta}{2}e^{-i\frac{\chi}{2}}(\downarrow)_{\theta,\chi}, \label{Zb}
             \end{align}
        \end{subequations}
        in which case the quantization direction of the basis vector rotates from the $z$ axis to the orientation ($\theta$, $\chi$). So the spin-triplet pairs have the following transformation:
         \begin{subequations}
            \begin{align}
            (\uparrow\uparrow)_{z}=&e^{i\chi}\left[\cos^2\frac{\theta}{2}(\uparrow\uparrow)_{\theta,\chi}+\sin^2\frac{\theta}{2}(\downarrow\downarrow)_{\theta,\chi}\right] \nonumber \\
                                   &-\sin\frac{\theta}{2}\cos\frac{\theta}{2}e^{i\chi}(\uparrow\downarrow+\downarrow\uparrow)_{\theta,\chi}, \label{Za}\\
            (\downarrow\downarrow)_{z}=&e^{-i\chi}\left[\sin^2\frac{\theta}{2}(\uparrow\uparrow)_{\theta,\chi}+\cos^2\frac{\theta}{2}(\downarrow\downarrow)_{\theta,\chi}\right] \nonumber \\
                                   &+\sin\frac{\theta}{2}\cos\frac{\theta}{2}e^{-i\chi}(\uparrow\downarrow+\downarrow\uparrow)_{\theta,\chi}. \label{Zb}
           \end{align}
         \end{subequations}

       \begin{figure*}[ptb]
           \centering
           \includegraphics[width=5.0in]{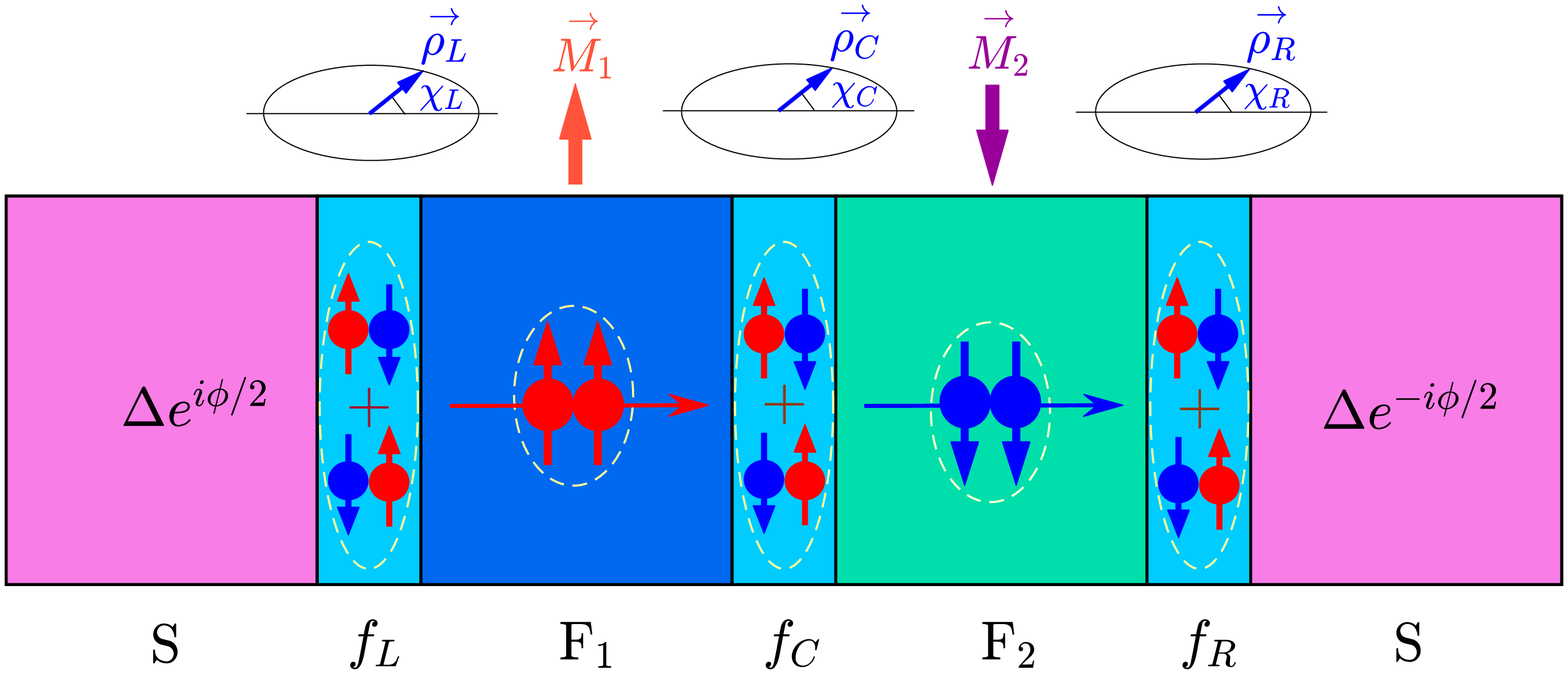}
           \caption{Transformation process of the spin-triplet pairs in the S/$f_L$-F$_1$-$f_C$-F$_2$-$f_R$/S junction. The F$_1$ and F$_2$ are strongly spin-polarized half-metals with antiparallel magnetizations $\mathop{M_1}\limits^{\rightarrow}$ and $\mathop{M_2}\limits^{\rightarrow}$, and the interfacial barriers $f_L$, $f_C$, and $f_R$ have misaligned magnetizations $\mathop{\rho_{L}}\limits^{\rightarrow}$, $\mathop{\rho_C}\limits^{\rightarrow}$, and $\mathop{\rho_R}\limits^{\rightarrow}$. The azimuthal angles of $\mathop{\rho_{L}}\limits^{\rightarrow}$, $\mathop{\rho_C}\limits^{\rightarrow}$, and $\mathop{\rho_R}\limits^{\rightarrow}$ are denoted by $\chi_L$, $\chi_C$, and $\chi_R$, respectively. The spin-triplet pairs ($\uparrow\downarrow+\downarrow\uparrow$) are usually generated in the $f_L$, $f_C$, and $f_R$ regions, which can be converted to the equal-spin triplet pairs ($\uparrow\uparrow$) in the F$_1$ layer or ($\downarrow\downarrow$) in the F$_2$ layer. When the spin-triplet pairs transport from the left $f_L$ region to the right $f_R$ region, they will acquire an additional phase $2\chi_C-\chi_L-\chi_R$.}
           \label{Fig5}
       \end{figure*}

      The transport process of the Cooper pairs in the S/$f_L$-F$_1$-$f_C$-F$_2$-$f_R$/S junction can be divided into the following steps:

      (i) It is well known that, since the spin-singlet pairs $(\uparrow\downarrow-\downarrow\uparrow)$ penetrate from the left superconductor into the $f_L$ interfacial region, it induces a mixture of the spin-singlet pairs $(\uparrow\downarrow-\downarrow\uparrow)_{\theta_{L}, \chi_{L}}$ and the spin-triplet pairs $(\uparrow\downarrow+\downarrow\uparrow)_{\theta_{L}, \chi_{L}}$ in the $f_L$ region~\cite{Eschg}. When the magnetization direction of the F$_1$ layer is different from that in the $f_L$ interface, the spin-singlet pairs $(\uparrow\downarrow-\downarrow\uparrow)_{\theta_{L}, \chi_{L}}$, which are rotationally invariant, cannot survive in the F$_1$ layer due to the strong exchange field. But the spin-triplet pairs $(\uparrow\downarrow+\downarrow\uparrow)_{\theta_{L}, \chi_{L}}$ can be transformed into a combination of the equal-spin-triplet pairs and the opposite-spin-triplet pairs when viewed with respect to the $z$ axis, see formula~(\ref{FLb}). Because the magnetization in the interfaces lies in the $x$-$y$ plane ($\theta_L=\theta_C=\theta_R=\pi/2$), the spin-triplet pairs $(\uparrow\downarrow+\downarrow\uparrow)_{z}$ disappear due to former factor $\cos\theta=0$. Moreover, the F$_1$ layer is defined as being polarized along the $z$ axis. So the spin-triplet pairs $(\uparrow\uparrow)_z$ can survive in the F$_1$ layer, but the spin-triplet pairs $(\downarrow\downarrow)_z$ are not allowed. When the spin-triplet pairs transfer from the $f_L$ interface into the F$_1$ layer, a conversion process can be obtained:
     \begin{equation}
            (\uparrow\downarrow+\downarrow\uparrow)_{\theta_L,\chi_L}\longrightarrow-e^{-i\chi_L}(\uparrow\uparrow)_{z}. \label{FLHM1}
     \end{equation}

      (ii) When the spin-triplet pairs pass from the F$_1$ layer into the $f_C$ interface, they have a transition process
         \begin{align}
            (\uparrow\uparrow)_{z}=&\frac{1}{2}e^{i\chi_C}\left[(\uparrow\uparrow)_{\theta_C,\chi_C}+(\downarrow\downarrow)_{\theta_C,\chi_C}\right] \nonumber \\
                                   &-\frac{1}{2}e^{i\chi_C}(\uparrow\downarrow+\downarrow\uparrow)_{\theta_C,\chi_C}. \label{HM1FC}
         \end{align}
         If one ignores the contribution of the first term on the right hand of Eq.~(\ref{HM1FC}), a simplified process can be obtained
         \begin{equation}
             (\uparrow\uparrow)_{z}\longrightarrow-\frac{1}{2}e^{i\chi_C}(\uparrow\downarrow+\downarrow\uparrow)_{\theta_C,\chi_C}.
             \label{HM1FC_2}
         \end{equation}

      (iii) Following the above process, the spin-triplet pairs transport from the $f_C$ interface into the F$_2$ layer, in which case they undergo a transformation
         \begin{equation}
            (\uparrow\downarrow+\downarrow\uparrow)_{\theta_C,\chi_C}=-e^{-i\chi_C}(\uparrow\uparrow)_z+e^{i\chi_C}(\downarrow\downarrow)_z.
             \label{FCHM1}
         \end{equation}
      Because the magnetization of the F$_2$ layer is antiparallel to the $z$ axis, the spin-triplet pairs $(\uparrow\uparrow)_z$ will be completely suppressed. As a result, we have a transformation process:
         \begin{equation}
            (\uparrow\downarrow+\downarrow\uparrow)_{\theta_C,\chi_C}\longrightarrow{e^{i\chi_C}(\downarrow\downarrow)_z}.
             \label{FCHM2}
         \end{equation}

      (iv) When the spin-triplet pairs move from the F$_2$ layer to the $f_R$ interface, they experience a transformation process:
         \begin{align}
            (\downarrow\downarrow)_{z}=&\frac{1}{2}e^{-i\chi_R}\left[(\uparrow\uparrow)_{\theta_R,\chi_R}+(\downarrow\downarrow)_{\theta_R,\chi_R}\right] \nonumber \\
            &+\frac{1}{2}e^{-i\chi_R}(\uparrow\downarrow+\downarrow\uparrow)_{\theta_R,\chi_R}. \label{FR1}
         \end{align}

      If the first term on the right hand of Eq.~(\ref{FR1}) is omitted, the following result can be achieved
      \begin{equation}
            (\downarrow\downarrow)_{z}\longrightarrow\frac{1}{2}e^{-i\chi_R}(\uparrow\downarrow+\downarrow\uparrow)_{\theta_R,\chi_R}. \label{FR2}
      \end{equation}
      The transmission process of the spin-triplet pairs through the entire S/$f_L$-F$_1$-$f_C$-F$_2$-$f_R$/S junction can be summarized as
      \begin{align}
            &(\uparrow\downarrow+\downarrow\uparrow)_{\theta_L,\chi_L}\xrightarrow{f_L\Rightarrow \rm{F_1}}-e^{-i\chi_L}(\uparrow\uparrow)_{z}\xrightarrow{\rm{F_1}\Rightarrow\emph{f}_\emph{C}} \nonumber \\
            &e^{i(\chi_C-\chi_L)}(\uparrow\downarrow+\downarrow\uparrow)_{\theta_C,\chi_C} \xrightarrow{f_C\Rightarrow \rm{F_2}}e^{i(2\chi_C-\chi_L)}(\downarrow\downarrow)_{z} \nonumber \\
            &\xrightarrow{\rm{F_2}\Rightarrow\emph{f}_\emph{R}}e^{i(2\chi_C-\chi_L-\chi_R)}(\uparrow\downarrow+\downarrow\uparrow)_{\theta_R,\chi_R}
             \label{FLFCFR},
      \end{align}
      where we discard the factors in front of the spin-triplet pairs. The above process is illustrated visually in Fig.~\ref{Fig5}. The spin-triplet pairs acquire an additional phase $2\chi_C-\chi_L-\chi_R$ when they pass through the entire system. The obtained phase may directly enter into the current phase to produce an expression $I^{\rm{3D}}(\phi)=I^{\rm{3D}}_{c}\sin(\phi+2\chi_C-\chi_L-\chi_R)$. This qualitative interpretation is consistent with the numerical results we obtained above.

      \begin{figure*}[ptb]
         \centering
         \includegraphics[width=7.0in]{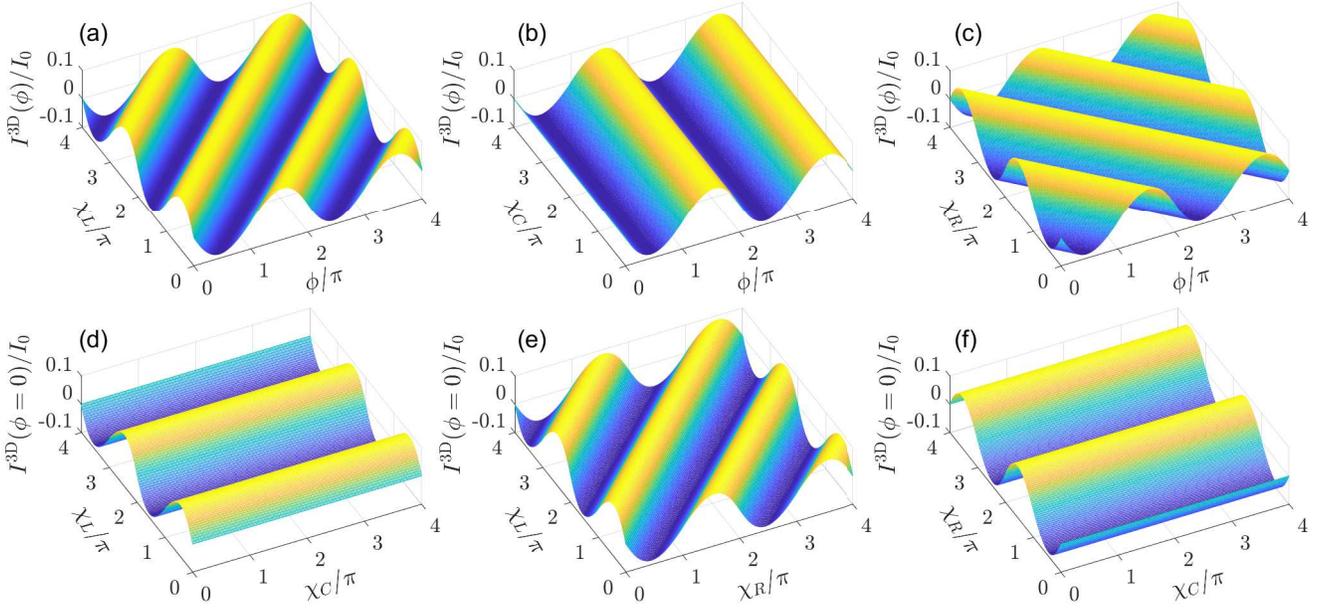}
         \caption{The top row illustrates the current-phase relation $I^{\rm{3D}}(\phi)$ as a function of the azimuthal angles $\chi_L$ (a), $\chi_C$ (b), and $\chi_R$ (c). The bottom row shows the spontaneous supercurrent $I^{\rm{3D}}(\phi=0)$ as a function of ($\chi_L$, $\chi_C$) (d), ($\chi_L$, $\chi_R$) (e), and ($\chi_R$, $\chi_C$) (f). The unlabeled azimuthal angles in each panel take the value of 0, and the other parameters are $h/E_F=1.05$, $k_Fd=50$, $P_L=P_R=1$, and $P_C=1.5$. The results shown are for the S/$f_L$-F$_{1}$-$f_C$-F$_{2}$-$f_R$/S junction with parallel magnetizations.}
         \label{Fig6}
      \end{figure*}

      \subsection{The Josephson current in the S/$f_L$-F$_{1}$-$f_C$-F$_{2}$-$f_R$/S junction with parallel magnetization configurations}

      In the following we investigate the contribution of the interfacial azimuthal angles to the Josephson current when the F$_1$ and F$_2$ layers are parallel half-metals ($h_{1}=h_{2}=h$ and $h/E_F=1.05$). The top row of Fig.~\ref{Fig6} shows the dependence of the current-phase relation $I^{\rm{3D}}(\phi)$ on the azimuthal angles. As seen, $I^{\rm{3D}}(\phi)$ oscillates with $\chi_L$ and $\chi_R$ but does not vary with $\chi_C$. The current magnitude is reduced compared to that in the junction with the antiparallel magnetizations. Moreover, the bottom row of Fig.~\ref{Fig6} presents the influence of the azimuthal angles on the spontaneous supercurrent $I^{\rm{3D}}(\phi=0)$. It is observed that $\chi_C$ has no contribution to $I^{\rm{3D}}(\phi=0)$. The variation characteristic of $I^{\rm{3D}}(\phi=0)$ with $\chi_L$ is different from that in the antiparallel S/$f_L$-F$_{1}$-$f_C$-F$_{2}$-$f_R$/S junction. From the dependence of the current on the phase difference and the azimuthal angles, one can deduce a current-phase relation $I^{\rm{3D}}(\phi)=I^{\rm{3D}}_{c}\sin(\phi+\chi_R-\chi_L+\pi)$.

      The physical picture leading to this relation can be explained by the transport process of the spin-triplet pairs:
      \begin{align}
            &(\uparrow\downarrow+\downarrow\uparrow)_{\theta_L,\chi_L}\xrightarrow{f_L\Rightarrow \rm{F_1}}-e^{-i\chi_L}(\uparrow\uparrow)_{z}\xrightarrow{\rm{F_1}\Rightarrow\emph{f}_\emph{C}} \nonumber \\
            &e^{i(\chi_C-\chi_L)}(\uparrow\downarrow+\downarrow\uparrow)_{\theta_C,\chi_C} \xrightarrow{f_C\Rightarrow \rm{F_2}}-e^{-i\chi_L}(\uparrow\uparrow)_{z} \nonumber \\
            &\xrightarrow{\rm{F_2}\Rightarrow\emph{f}_\emph{R}}e^{i(\chi_R-\chi_L)}(\uparrow\downarrow+\downarrow\uparrow)_{\theta_R,\chi_R}
             \label{FLFCFR}.
      \end{align}
      In the above process, when the spin-triplet pairs pass from the F$_1$ layer into the $f_C$ interface, they transform from $(\uparrow\uparrow)_z$ to $(\uparrow\downarrow+\downarrow\uparrow)_{\theta_C,\chi_C}$ and acquire an additional phase $\chi_C$. Continually, if $(\uparrow\downarrow+\downarrow\uparrow)_{\theta_C,\chi_C}$ penetrate from the $f_C$ interface into the F$_2$ layer, they are converted into $(\uparrow\uparrow)_z$ and get another phase $-\chi_C$. The two phases resulting from the spin-triplet pairs crossing the $f_C$ interface can superimpose and cancel each other out. During the entire transmission process, the obtained phase is only related to the azimuthal angles $\chi_L$ and $\chi_R$. So we can say that the $f_C$ interface works as a conventional potential barrier and does not contribute any additional phase to the Josephson current. The above current characteristics are the same as those in the S/$f_L$-F-$f_R$/S junction, which we further demonstrate below.

      \begin{figure*}[ptb]
         \centering
         \includegraphics[width=7.0in]{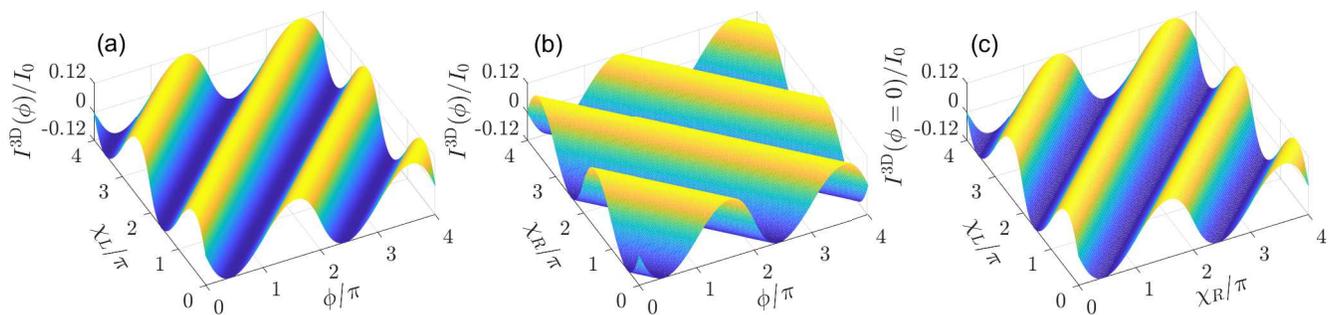}
         \caption{Current-phase relation $I^{\rm{3D}}(\phi)$ is shown as a function of the azimuthal angles $\chi_L$ (a) and $\chi_R$ (b). The spontaneous supercurrent $I^{\rm{3D}}(\phi=0)$ is shown as a function of the azimuthal angles ($\chi_L$, $\chi_R$) (c). The unlabeled azimuthal angles in each panel are taken as 0, and the other parameters are set to $h_1/E_F=1.05$, $k_Fd_1=50$, and $P_L=P_R=1$. The results shown are for the S/$f_L$-F-$f_R$/S junction ($k_Fd_2=0$ and $P_C=0$).}
         \label{Fig7}
      \end{figure*}

      \subsection{The Josephson current in the S/$f_L$-F-$f_R$/S junction}

      In Figs.~\ref{Fig7}(a) and \ref{Fig7}(b) we illustrate the variation of the current-phase relation $I^{\rm{3D}}(\phi)$ with the azimuthal angles $\chi_L$ and $\chi_R$ in the S/$f_L$-F-$f_R$/S junction. It can be seen that, except for the increased amplitude, $I^{\rm{3D}}(\phi)$ has the same characteristics as shown in Figs.~\ref{Fig6}(a) and \ref{Fig6}(c). Moreover, Fig.~\ref{Fig7}(c) shows the spontaneous current $I^{\rm{3D}}(\phi=0)$ as a function of the azimuthal angles $\chi_L$ and $\chi_R$, which is the same as that shown in Fig.~\ref{Fig6}(e). These behaviors demonstrate that the Josephson current in the S/$f_L$-F-$f_R$/S junction satisfies a relation $I^{\rm{3D}}(\phi)=I^{\rm{3D}}_{c}\sin(\phi+\chi_R-\chi_L+\pi)$, which is consistent with the results in Refs.~\cite{Eschrig,VBYVNa,RGrein}. This relation can be explained by the transport process of the spin-triplet pairs:
      \begin{align}
         &(\uparrow\downarrow+\downarrow\uparrow)_{\theta_L,\chi_L}\xrightarrow{f_L\Rightarrow \rm{F}}-e^{-i\chi_L}(\uparrow\uparrow)_{z}\xrightarrow{\rm{F}\Rightarrow\emph{f}_\emph{R}} \nonumber \\
         &e^{i(\chi_R-\chi_L)}(\uparrow\downarrow+\downarrow\uparrow)_{\theta_R,\chi_R}
         \label{SFLHMFRS}.
      \end{align}
      Therefore, the same current-phase relation can be generated in the S/$f_L$-F-$f_R$/S junction and the parallel S/$f_L$-F$_1$-$f_C$-F$_2$-$f_R$/S junction. The last which must be noted is that the current magnitude in the S/$f_L$-F-$f_R$/S junction is almost the same as that in the antiparallel S/$f_L$-F$_1$-$f_C$-F$_2$-$f_R$/S junction (see Figs.~\ref{Fig4} and \ref{Fig7}). It indicates that adding the antiparallel F$_2$ layer and the $f_C$ interface does not significantly suppress the Josephson current.

     \section{Conclusion} \label{Sec4}

     On the basis of the exact numerical solution of the Bogoliubov-de Gennes equations, we have investigated the Josephson current in the S/$f_L$-F$_{1}$-$f_C$-F$_{2}$-$f_R$/S junction. If the magnetizations of the F$_1$ and F$_2$ layers are antiparallel, the critical current oscillates with the exchange field and the thickness of the F$_1$ and F$_2$ layers when the azimuthal angle $\chi_L$ takes some specific values. The oscillation is not significant for the other $\chi_L$. By contrast, when the magnetization at the $f_C$ interface is perpendicular to that at the $f_L$ and $f_R$ interfaces, the critical current reaches a larger value and is rarely affected by the exchange field and the thickness. For other directions of the $f_C$ interface, the critical current decreases and a significant oscillation effect occurs. Interestingly, the critical current will no longer change with the azimuthal angles and the ferromagnetic thickness once the antiparallel ferromagnets increase up to the half-metallic phase. In this situation, the Josephson current will gain an additional phase $\phi_0$ to form an anomalous supercurrent $I^{\rm{3D}}(\phi)=I^{\rm{3D}}_{c}\sin(\phi+\phi_0)$ with $\phi_0=2\chi_C-\chi_L-\chi_R$. This feature reveals direct coupling between the interface magnetizations and the Josephson phase difference. We attribute this anomalous effect to the phase superposition. The spin-triplet pairs capture a phase as they pass through each interface. All the captured phases are superposed to contribute to the Josephson current. When the F$_1$ and F$_2$ layers become parallel half-metals, the central $f_C$ interface cannot provide any phase to the Josephson current and $\phi_0=\chi_R-\chi_L+\pi$. The $f_C$ interface works like a conventional potential barrier because the spin-triplet pairs passing through the $f_C$ interface create a phase cancellation effect. In such a case, the current-phase relation is the same as that in the S/$f_L$-F-$f_R$/S junction. The results we obtained above might be used in experiments to construct novel structures to manufacture the $\phi_0$-junction, and ultimately achieve the purpose of controlling the superconducting phase in superconducting spintronics.

     \section*{Acknowledgments}

     The authors thank A. I. Buzdin for useful discussions and suggestions. This work was supported by the National Natural Science Foundation of China (Grants No.12174238 and No.11604195), the Special Foundation for Theoretical Physics Research Program of China (Grant No.11747035), the Natural Science Basic Research Program of Shaanxi (Programs No.2020JM-597 and No.2021JQ-748), and the Scientific Research Foundation of Shaanxi University of Technology (Grant No.SLGKY2006). J. Wu was supported by Natural Science Foundation of Guangdong Province (Grant No.2017B030308003 and No.2019B121203002), Guangdong Innovative and Entrepreneurial Research Team Program (No.2016ZT06D348) and Science, Technology, and Innovation Commission of Shenzhen Municipality (Grants No.KYTDPT20181011104202253 and No. JCYJ20170412152620376).

    \end{document}



   \title{Supplementary material for ``Anomalous supercurrent modulated by interfacial magnetizations in Josephson junctions with ferromagnetic bilayers''}

   \author{Hao Meng}
   \affiliation{Shenzhen Institute for Quantum Science and Engineering (SIQSE), Southern University of Science and Technology, Shenzhen 518055, China}
   \affiliation{School of Physics and Telecommunication Engineering, Shaanxi University of Technology, Hanzhong 723001, China}
   \affiliation{International Quantum Academy, Shenzhen, 518048, China}
   \author{Xiuqiang Wu}
   \affiliation{Department of Physics, Yancheng Institute of Technology, Yancheng 224051, China}
   \author{Yajie Ren}
   \affiliation{School of Physics and Telecommunication Engineering, Shaanxi University of Technology, Hanzhong 723001, China}
   \author{Jiansheng Wu}
   \email{wujs@sustech.edu.cn}
   \affiliation{Shenzhen Institute for Quantum Science and Engineering (SIQSE), Southern University of Science and Technology, Shenzhen 518055, China}
   \affiliation{International Quantum Academy, Shenzhen, 518048, China}

   \date{\today}

   \maketitle

   \section{introduction for experimental implementation}

   Initially, the Birge group~\cite{TSKhaire,MAKhasa} observed a long-range supercurrent in Josephson junctions of the form S/$f_L$-Co-Ru-Co-$f_R$/S, where the central Co-Ru-Co trilayer was a synthetic antiferromagnet (SAF), and the interfacial $f_L$ and $f_R$ layers were PdNi or CuNi weakly ferromagnetic alloys. The thin Ru layer induces antiparallel exchange coupling between the domains in the two Co layers, leaving nearly zero net magnetization in the junctions. The purpose of using the SAF is to cancel the magnetic flux produced by the Co domain structure. Shortly thereafter, they use thin Ni layers as the interfacial $f_{L}$ and $f_{R}$ layers. After applying a large in-plane field to the junction, the SAF undergoes a ``spin-flop'' transition whereby the two Co layers end up with their magnetizations perpendicular to the applied field direction, while the $f_{L}$ and $f_{R}$ layers are magnetized parallel to the applied field~\cite{CaroKK}. This configuration with perpendicular magnetizations optimizes the magnitude of the spin-triplet supercurrent and resulted in a factor of 20 enhancement of supercurrent in the junctions compared to the as-grown state.

   To modulate the Josephson current experimentally, an external magnetic field can be applied to rotate the magnetization direction of the interfaces. Recently, Banerjee \emph{et al.}~\cite{NBanJWARob} observed that the amplitude of spin-triplet supercurrent in the S/$f_{L}$-F-$f_{R}$/S junction can be controlled by applying the magnetic field to manipulate the magnetization direction of the $f_{L}$ and $f_{R}$ interfaces. Similarly, for the antiparallel S/$f_{L}$-F$_{1}$-$f_{C}$-F$_{2}$-$f_{R}$/S junction, the long-range triplet supercurrent can be controllably turned on and off by rotating the magnetization of the $f_{R}$ interface~\cite{WillMMWP}. Moreover, the ground-state phase across the antiparallel S/$f_{L}$-F$_{1}$-$f_{C}$-F$_{2}$-$f_{R}$/S junction can be toggled between 0 and $\pi$ by reversing the magnetization direction of one of the $f_{L}$ and $f_{R}$ interfaces~\cite{JAGVAgui}. The above experimental observations on the Josephson current manipulation may be used in spintronic devices in the future.

    Compared with the S/$f_{L}$-F-$f_{R}$/S junction, the antiparallel S/$f_{L}$-F$_{1}$-$f_{C}$-F$_{2}$-$f_{R}$/S junction has the following advantages for experimental observations: (i) Since the central F$_{1}$-$f_{C}$-F$_{2}$ trilayer is a SAF Co-Ru-Co, the magnetizations of the two Co layers exchange-coupled antiparallel to each other via the Ru layer. The advantage of using a SAF rather than just a single F layer is to cancel the magnetic flux produced by the Co domain structure inside the junction. Samples containing the SAF instead of only a single Co layer exhibit nearly perfect `Fraunhofer patterns' when the critical current is measured as a function of a magnetic field applied perpendicular to the current direction~\cite{MAKhasa}. (ii) According to our calculated results, the additional phase $\phi_{0}$ obtained by supercurrent in the S/$f_{L}$-F-$f_{R}$/S junction is related to the angle difference $\chi_{R}-\chi_{L}$. Because the two angles ($\chi_{L}$ and $\chi_{R}$) have opposite signs, the contribution of rotating angle $\chi_{L}$ to the phase $\phi_{0}$ is different from rotating angle $\chi_{R}$. Moreover, if the angles $\chi_{L}$ and $\chi_{R}$ are rotated by the same degree at the same time, there is no contribution to the phase $\phi_0$. In contrast, the phase $\phi_0$ in the antiparallel S/$f_{L}$-F$_{1}$-$f_{C}$-F$_{2}$-$f_{R}$/S junction is proportional to $-\chi_{L}-\chi_{R}$. In such a case, rotating angle $\chi_{L}$ produces the same effect as rotating angle $\chi_{R}$. If the angles $\chi_{L}$ and $\chi_{R}$ are rotated by the same degree at the same time, the effect on the phase $\phi_{0}$ not only does not disappear but also produces a superimposed contribution. (iii) In the antiparallel S/$f_{L}$-F$_{1}$-$f_{C}$-F$_{2}$-$f_{R}$/S junction, the angle $\chi_{C}$ at the $f_{C}$ interface can be used as an additional parameter to modulate the phase $\phi_{0}$, which satisfies a relationship $\phi_{0}\propto2\chi_{C}$. Its physical meaning is that the phase $\phi_{0}$ obtains a $\pi$ phase shift as long as the $f_{C}$ interface rotates by $\pi/2$. Correspondingly, to shift the phase $\phi_{0}$ by $\pi$ in the original S/$f_{L}$-F-$f_{R}$/S junction, it is necessary to reverse the magnetization direction of the $f_{L}$ or $f_{R}$ interfaces. So the S/$f_L$-F$_1$-$f_C$-F$_2$-$f_R$/S structure we considered may pave the way for interesting implementations of the $\phi_{0}$ junctions in superconducting spintronics.

    \begin{figure*}[ptb]
      \centering
      \includegraphics[width=6.2in]{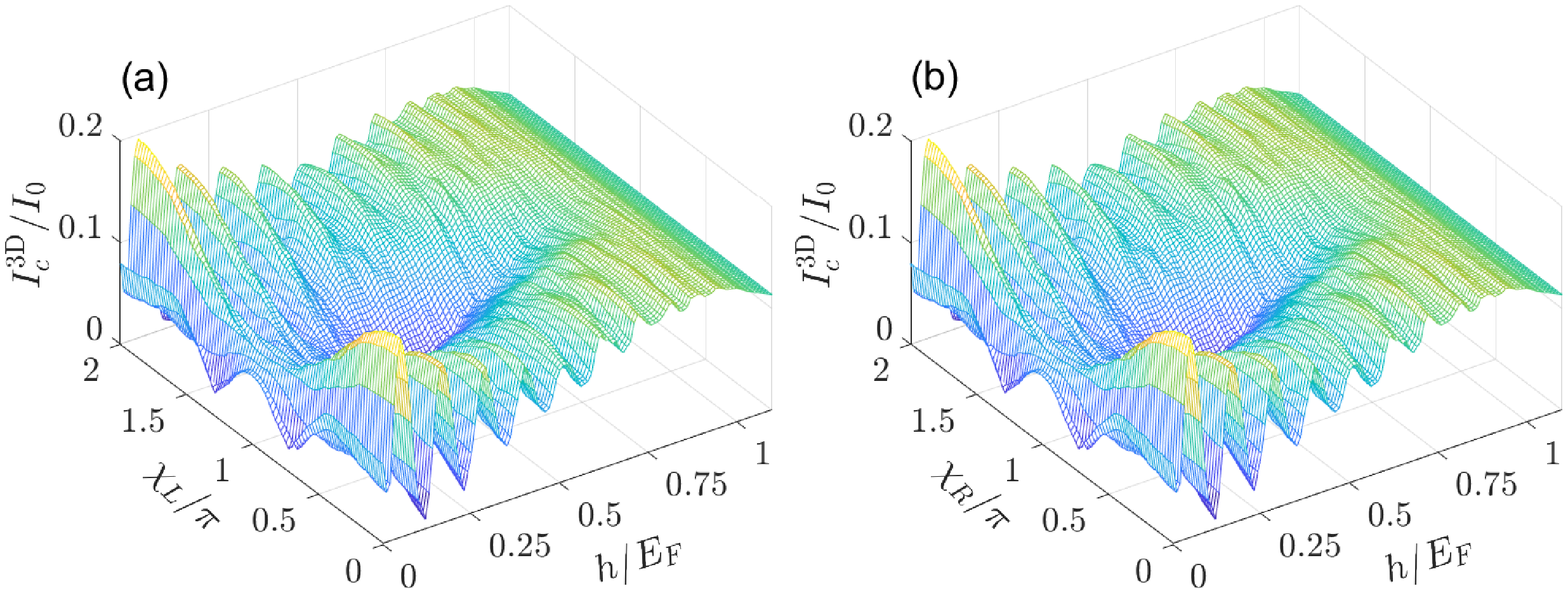}
      \caption{(a) The critical current $I_c^{\rm{3D}}$ versus the azimuthal angle $\chi_{L}$ and the exchange field $h/E_F$ for $\chi_{C}=\chi_{R}=0$. (b) $I_{c}^{3\rm{D}}$ versus $\chi_{R}$ and $h/E_F$ for $\chi_{L}=\chi_{C}=0$. Here we choose the ferromagnetic thickness $k_Fd=50$ and the spin-active barriers $P_{L}=P_{R}=1$ and $P_{C}=1.5$.}
      \label{Figs1}
    \end{figure*}

    In the actual measurement device, the largest spin-triplet supercurrent can be observed when the S/$f_{L}$-F$_{1}$-$f_{C}$-F$_{2}$-$f_{R}$/S junction is optimally constructed. According to recent experimental results~\cite{TSKhaire,MAKhasa,CaroKK}, the F$_{1}$ and F$_{2}$  layers could be taken as Co with 20-nm-thick. The $f_{L}$ and $f_{R}$ interfaces are implemented as a 4-nm-thick layer of PdNi alloy or a 1.5-nm-thick Ni layer, and the $f_{C}$ interface is a 0.6-nm-thick Ru layer. Furthermore, to observe the spontaneous supercurrent more obviously, one can replace ferromagnets Co with half-metals CrO$_{2}$~\cite{RSKeizerSTB,MSAFCzeschka,DSManzano,MEgilmez}.

    \section{the dependence of the Josephson current on interface parameters}

    In the following figures, we plot the variation of the Josephson current with interface parameters in the S/$f_L$-F$_{1}$-$f_C$-F$_{2}$-$f_R$/S junction with antiparallel magnetizations.

    Figure~\ref{Figs1} shows the dependence of the critical current $I^{\rm{3D}}_c$ on the azimuthal angles and the exchange field. As illustrated in Fig.~\ref{Figs1}(a), $I_{c}^{3\rm{D}}$ exhibits an oscillatory characteristic with increasing $h/E_F$. The current amplitude first decreases and then increases, and its oscillation completely disappears as the ferromagnets transform into half-metals ($h/E_F\geq1$). Furthermore, the oscillating effect of $I_{c}^{3\rm{D}}$ is more prominent at $\chi_L=0$ and $2\pi$ but almost no longer appears at $\chi_L=\pi$. On the other hand, the variation of $I_{c}^{3\rm{D}}$ with $\chi_{L}$ shows different characteristics in different ranges of $h/E_F$. In some weak field ranges ($0\leq{h/E_F}\leq0.02$ and $0.09\leq{h/E_F}\leq0.13$), $I_{c}^{3\rm{D}}$ has a larger value at $\chi_{L}=\pi$, but in the moderate and strong field ranges ($0.25\leq{h/E_F}\leq0.95$), $I_{c}^{3\rm{D}}$ becomes minimum at $\chi_{L}=\pi$. Interestingly enough, when the exchange field reaches a critical value ($h/E_F\geq1$), $I_{c}^{3\rm{D}}$ maintains a constant value and does not change with $\chi_{L}$. It is worth emphasizing that the change of $I_{c}^{3\rm{D}}$ with $\chi_R$ is exactly the same as that with $\chi_L$ [see Fig.~\ref{Figs1}(b)]. From this feature, we can infer that the left and right interfaces play the same role in the transport process of the Josephson current.

    \begin{figure*}[ptb]
      \centering
      \includegraphics[width=5.8in]{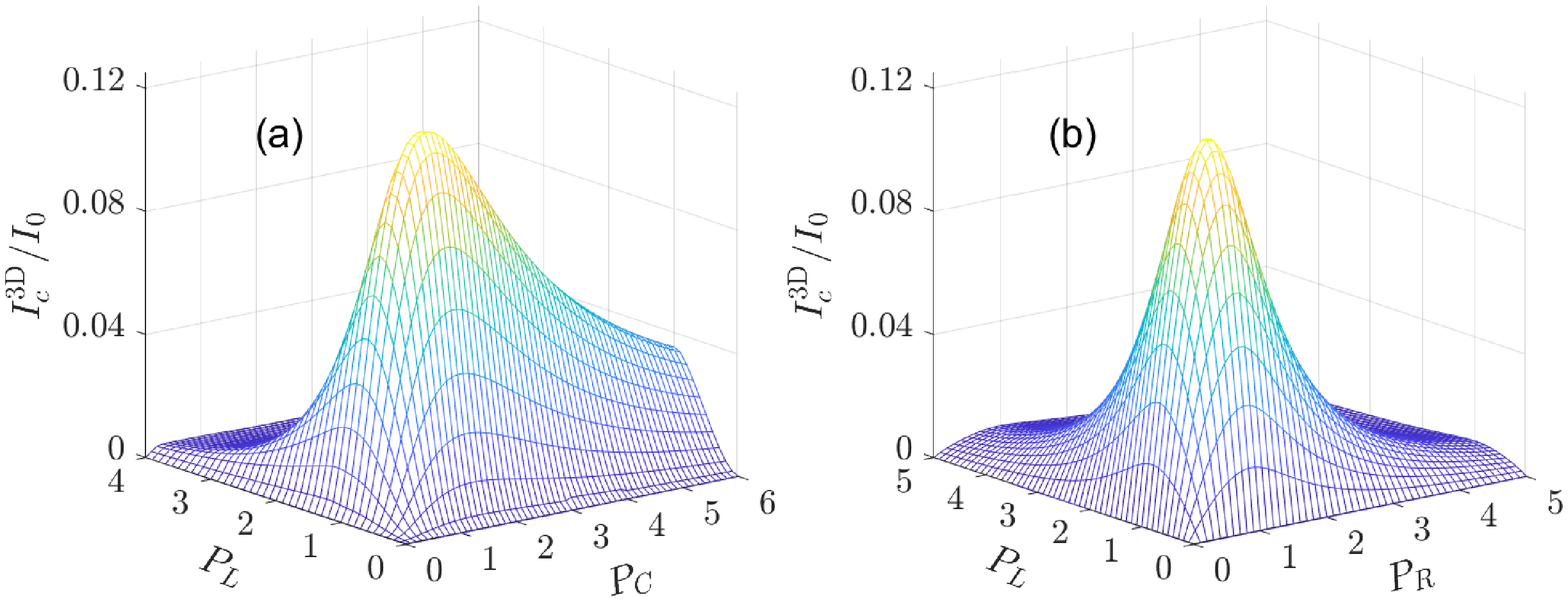}
      \caption{(a) The critical current $I_{c}^{3\rm{D}}$ versus the strength of the spin-active barriers $P_{L}$ and $P_{C}$ in the case of $P_{L}=P_{R}$. (b) $I_{c}^{3\rm{D}}$ versus $P_{L}$ and $P_{R}$ for $P_{C}=1.5$. Here we choose $h/E_{F}=1.05$, $k_Fd=50$, $\chi_{L}=\chi_{R}=0$, and $\chi_C=\pi/2$.}
      \label{Figs2}
    \end{figure*}

    \begin{figure*}[ptb]
      \centering
      \includegraphics[width=7.2in]{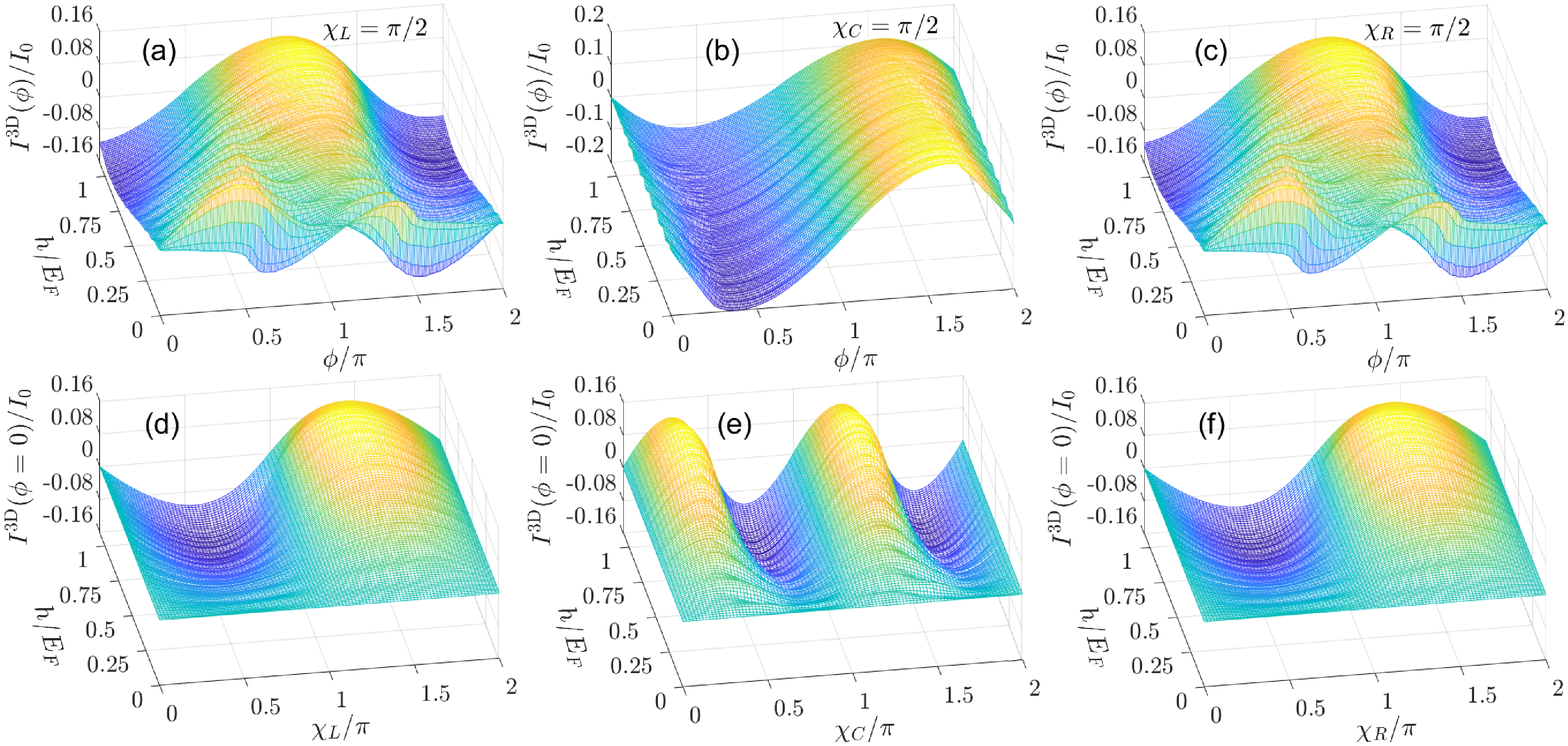}
      \caption{Current-phase relation $I^{\rm{3D}}(\phi)$ versus exchange field $h/E_F$ for $\chi_L=\pi/2$ (a), $\chi_C=\pi/2$ (b), and $\chi_R=\pi/2$ (c). Here the unlabeled azimuthal angles in panels (a)--(c) are set to 0. Spontaneous supercurrent $I^{\rm{3D}}(\phi=0)$ versus $h/E_F$ and $\chi_L$ for $\chi_C=\chi_R=0$ (d), $I^{\rm{3D}}(\phi=0)$ versus $h/E_F$ and $\chi_C$ for $\chi_L=\chi_R=0$ (e), and $I^{\rm{3D}}(\phi=0)$ versus $h/E_F$ and $\chi_R$ for $\chi_L=\chi_C=0$ (f). In all panels, the parameters are chosen as $k_Fd=50$, $P_L=P_R=1$, and $P_C=1.5$.}
      \label{Figs3}
    \end{figure*}

    In Fig.~\ref{Figs2}, we present the dependence of $I^{\rm{3D}}_c$ on the strengths of the interfacial spin-active barriers when both the ferromagnets become half-metals. As seen in figure~\ref{Figs2}(a), without the spin-active barriers ($P_L=0$ or/and $P_C=0$), $I^{\rm{3D}}_c$ is zero. As the interfacial barriers increase, $I^{\rm{3D}}_c$ increases rapidly and reaches a maximum for $P_L=1$ and $P_C=1.5$, then decreases at larger values of $P_L$ and $P_C$. It is interesting to note that $I^{\rm{3D}}_c$ decreases faster with $P_L$ but slower with $P_C$. By contrast, $I^{\rm{3D}}_c$ displays the same dependence on $P_L$ and $P_R$ and reaches a maximum at $P_L=P_R=1$ [see Fig.~\ref{Figs2}(b)]. There are two specific cases to be noted. First, when the left and right interfacial barriers do not exist ($P_L=P_R=0$), no matter what the value of the central interfacial barrier $P_C$ takes, $I^{\rm{3D}}_c$ is always absent. This situation is different from what we have previously discussed in Ref.~\cite{HMBuzdin}. It was found that, for the S/F$_1$-F$_2$/S junction with antiparallel magnetic moments, the spin-independent potential barrier at the F$_1$/F$_2$ interface can result in a large oscillation of the critical current with the ferromagnetic thickness or exchange field, which can be attributed to the resonant tunneling of the spin-singlet pairs. In our present case, because the F$_1$ and F$_2$ layers are turned into half-metals, the spin-singlet pairs are suppressed completely. Although the central $f_C$ interface works as a spin-active potential barrier, the equal-spin triplet pairs still cannot be induced in the system due to the absence of the $f_L$ and $f_R$ interfaces. Second, the absence of the central $f_C$ interface also leads to the disappearance of the Josephson current, which can be attributed to the fact that the spin-up triplet pairs ($\uparrow\uparrow$) in the F$_1$ layer cannot be converted into the spin-down triplet pairs ($\downarrow\downarrow$) in the F$_2$ layer. Recent experiments~\cite{TSKhaire,MAKhasa,CaroKK} have demonstrated the critical role of the central interface layer. In the S/$f_L$-Co-Ru-Co-$f_R$/S junction, the Ru layer acts as a bridge to maintain the spin-triplet supercurrent at a large value. Without the Ru layer, the critical current becomes very small. In addition, compared with our previous work~\cite{MengWuZheng}, the present results do not show the oscillation of the critical current with the strengths of the interfacial barriers ($P_L$ and $P_R$). That is because we simplify the $f_L$ and $f_R$ layers into the spin-active barriers, in which case the thicknesses of both layers are ignored.

    \begin{figure*}[ptb]
      \centering
      \includegraphics[width=6.2in]{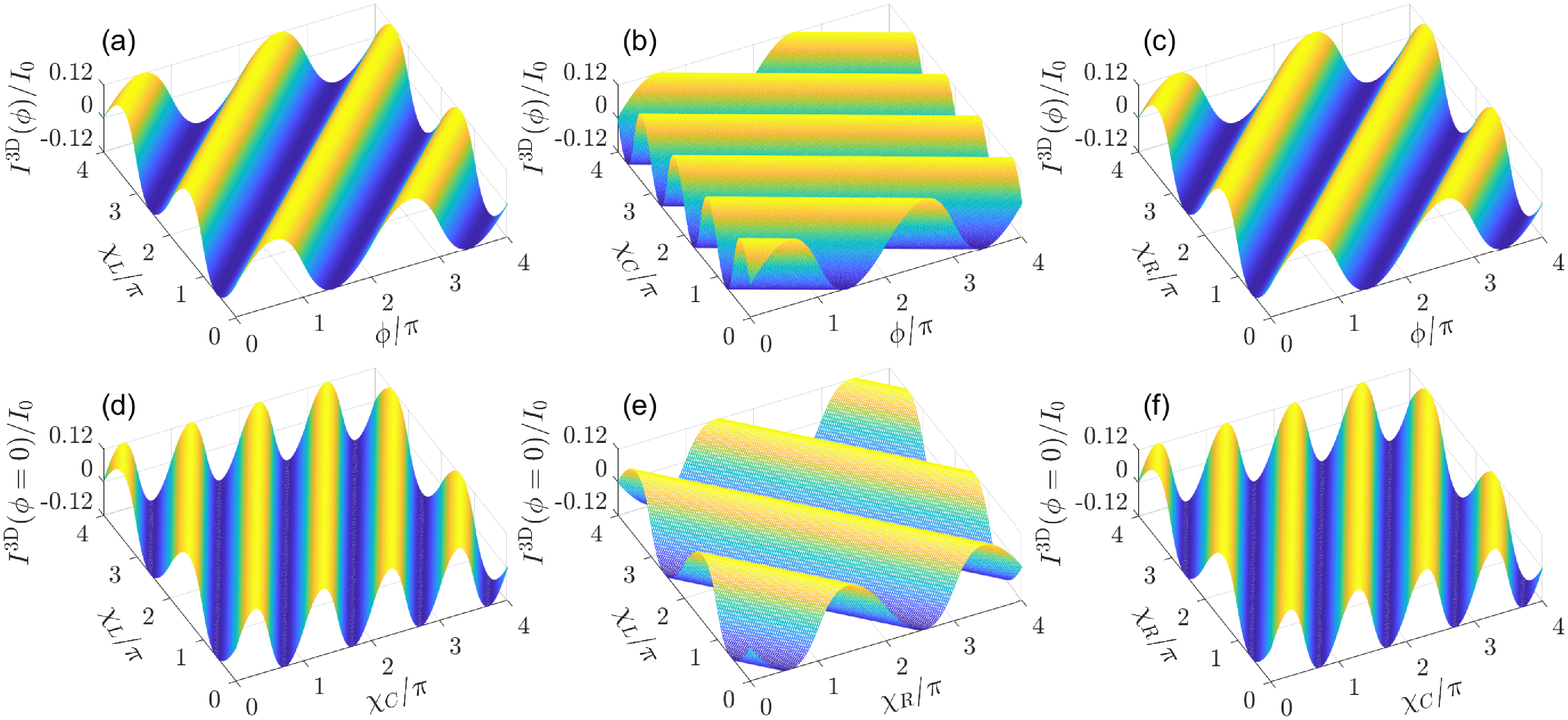}
      \caption{Current-phase relation $I^{\rm{3D}}(\phi)$ versus the azimuthal angles $\chi_L$ (a), $\chi_C$ (b), and $\chi_R$ (c). Spontaneous supercurrent $I^{\rm{3D}}(\phi=0)$ versus the azimuthal angles ($\chi_L$, $\chi_C$) (d), ($\chi_L$, $\chi_R$) (e), and ($\chi_R$, $\chi_C$) (f). The unlabeled azimuthal angles in each panel take the value of 0, and the other parameters are $h/E_F=1.05$, $k_Fd=50$, $P_L=P_R=1$, and $P_C=1.5$.}
      \label{Figs4}
    \end{figure*}

    The influence of the exchange field $h/E_F$ on the current-phase relation $I^{\rm{3D}}(\phi)$ and the spontaneous supercurrent $I^{\rm{3D}}(\phi=0)$ is illustrated in Fig.~\ref{Figs3}. We consider three special cases, where one interface is magnetized along the $y$ axis, and the other two interfaces are along the $x$ axis. (i) The magnetization of the $f_L$ interface is along the $y$ axis [see Fig.~\ref{Figs3}(a)]. Under a weak $h/E_F$, $I^{\rm{3D}}(\phi)$ cannot be represented as a sinusoidal function of the phase difference $\phi$. With the enhancement of $h/E_F$, the current amplitude is non-monotonic and accompanied by tiny oscillations. When $h/E_F$ reaches a high value, the oscillations disappear, and the current satisfies a relation $I^{\rm{3D}}(\phi)=I^{\rm{3D}}_c\sin(\phi-\pi/2)$ for $\chi_L=\pi/2$. (ii) The $f_C$ interface is oriented along the $y$ axis, as shown in Fig.~\ref{Figs3}(b). During the increase of $h/E_F$, the current amplitude has a slight oscillation and attenuation, and the current approximately maintains a sinusoidal relation. When both the ferromagnets are converted into half-metals ($h/E_F\geq1$), the current strictly follows a functional relation $I^{\rm{3D}}(\phi)=I^{\rm{3D}}_c\sin(\phi+\pi)$ for $\chi_C=\pi/2$. (iii) If the $f_R$ interface points to the $y$ axis, $I^{\rm{3D}}(\phi)$ shows the same characteristics as in case (i) [see Fig.~\ref{Figs3}(c)]. So we can get a similar relation $I^{\rm{3D}}(\phi)=I^{\rm{3D}}_c\sin(\phi-\pi/2)$ for $\chi_R=\pi/2$. In addition, as observed in Fig.~\ref{Figs3}(d), $I^{\rm{3D}}(\phi=0)$ does not exist at $h/E_F=0$, and its amplitude increases with increasing $h/E_F$. The amplitude of $I^{\rm{3D}}(\phi=0)$ reaches a maximum when the two ferromagnets are converted into half-metals. The spontaneous supercurrent satisfies a relation $I^{\rm{3D}}(\phi=0)=I^{\rm{3D}}_c\sin(-\chi_L)$ for $\chi_C=\chi_R=0$. In comparison, as shown in Fig.~\ref{Figs3}(e), the $I^{\rm{3D}}(\phi=0)$ amplitude still increases with $h/E_F$, but its oscillation period with $\chi_C$ will be halved. Therefore, the spontaneous supercurrent can be expressed as $I^{\rm{3D}}(\phi=0)=I^{\rm{3D}}_c\sin(2\chi_C)$ for $\chi_L=\chi_R=0$. Interestingly, the change of $I^{\rm{3D}}(\phi=0)$ with $\chi_R$ is the same as that with $\chi_L$ [see Fig.~\ref{Figs3}(f)]. So we can acquire a similar relation $I^{\rm{3D}}(\phi=0)=I^{\rm{3D}}_c\sin(-\chi_R)$ for $\chi_L=\chi_C=0$.

    In the following we derive the complete current-phase relation $I^{\rm{3D}}(\phi)$ for the entire system when the F$_1$ and F$_2$ layers become half-metals. The top row of Fig.~\ref{Figs4} illustrates the variation of the current-phase relation $I^{\rm{3D}}(\phi)$ with the azimuthal angles. As shown in Fig.~\ref{Figs4}(a), $I^{\rm{3D}}(\phi)$ oscillates with $\phi$ and $\chi_L$, and its oscillation period is both $2\pi$. The $I^{\rm{3D}}(\phi)$ curve shifts linearly to the right as $\chi_L$ increases. So the current has a relation $I^{\rm{3D}}(\phi)=I^{\rm{3D}}_c\sin(\phi-\chi_L)$ under the condition of $\chi_C=\chi_R=0$. In Fig.~\ref{Figs4}(b) the variation period of $I^{\rm{3D}}(\phi)$ with $\phi$ does not change, but its period with $\chi_C$ becomes $\pi$. The $I^{\rm{3D}}(\phi)$ curve shifts linearly to the left with increasing $\chi_C$. As a result, the current-phase relation can be expressed as $I^{\rm{3D}}(\phi)=I^{\rm{3D}}_c\sin(\phi+2\chi_C)$ for $\chi_L=\chi_R=0$. Moreover, $I^{\rm{3D}}(\phi)$ changes with $\chi_R$ in the same way as it does with $\chi_L$ [see Fig.~\ref{Figs4}(c)]. Accordingly, the current satisfies a relation $I^{\rm{3D}}(\phi)=I^{\rm{3D}}_c\sin(\phi-\chi_R)$ for $\chi_L=\chi_C=0$. The bottom row of Fig.~\ref{Figs4} shows the spontaneous supercurrent $I^{\rm{3D}}(\phi=0)$ as a function of the azimuthal angles. As seen in Fig.~\ref{Figs4}(d), the oscillation periods of $I^{\rm{3D}}(\phi=0)$ with $\chi_L$ and $\chi_C$ are $2\pi$ and $\pi$, respectively. The $I^{\rm{3D}}(\phi=0)$ curve shifts to the right as $\chi_L$ increases. Therefore, the current relation can be expressed as $I^{\rm{3D}}(\phi=0)=I^{\rm{3D}}_c\sin(2\chi_C-\chi_L)$ for $\chi_R=0$. On the other hand, $I^{\rm{3D}}(\phi=0)$ varies in the same way with $\chi_L$ and $\chi_R$, and its oscillation period is $2\pi$ [see Fig.~\ref{Figs4}(e)]. The $I^{\rm{3D}}(\phi=0)$ curve shifts to the left as $\chi_L$ increases. Thereupon, the current relation has a form $I^{\rm{3D}}(\phi=0)=I^{\rm{3D}}_c\sin(-\chi_L-\chi_R)$ for $\chi_C=0$. Furthermore, the curves of $I^{\rm{3D}}(\phi=0)$ with $\chi_R$ and $\chi_C$ are the same as that with $\chi_L$ and $\chi_C$ [see Fig.~\ref{Figs4}(f)]. So one can get a similar relation $I^{\rm{3D}}(\phi=0)=I^{\rm{3D}}_c\sin(2\chi_C-\chi_R)$ for $\chi_L=0$. Based on the above inferences, we obtain a final current-phase relation $I^{\rm{3D}}(\phi)=I^{\rm{3D}}_c\sin(\phi+2\chi_C-\chi_L-\chi_R)$ for the entire system.